\documentclass{aa}
\usepackage{txfonts}
\usepackage{graphicx}
\usepackage{natbib}
\usepackage{longtable}
\bibpunct{(}{)}{;}{a}{}{,}
\begin{document}
\title{Doppler factors, Lorentz factors and viewing angles for quasars, BL Lacertae objects and radio galaxies}

\author{T. Hovatta \inst{1} \and E. Valtaoja \inst{2,3} \and M. Tornikoski \inst{1} \and A. L\"ahteenm\"aki  \inst{1}}
\institute{Mets\"ahovi Radio Observatory, TKK, Helsinki University of Technology, Mets\"ahovintie 114, 02540 Kylm\"al\"a, Finland  \\ \email{tho@kurp.hut.fi} \and Tuorla Observatory, University of Turku, V\"ais\"al\"antie 20, 21500 Piikki\"o, Finland \and Department of Physics and Astronomy, University of Turku, Vesilinnantie 5, 20100 Turku, Finland}

\date{Received / Accepted}
\abstract
{}
{We have calculated variability Doppler boosting factors, Lorentz factors, and viewing angles for a large sample of sources by using total flux density observations at 22 and 37\,GHz and VLBI data.}
{We decomposed the flux curves into exponential flares and determined the variability brightness temperatures of the fastest flares. By assuming the same intrinsic brightness temperature for each source, we calculated the Doppler boosting factors for 87 sources. In addition we used new apparent jet speed data to calculate the Lorentz factors and viewing angles for 67 sources.}
{We find that all quasars in our sample are Doppler-boosted and that the Doppler boosting factors of BL Lacertae objects are lower than of quasars. The new Lorentz factors are about twice as high as in earlier studies, which is mainly due to higher apparent speeds in our analyses. The jets of BL Lacertae objects are slower than of quasars. There are some extreme sources with very high derived Lorentz factors of the order of a hundred. These high Lorentz factors could be real. It is also possible that the sources exhibit such rapid flares that the fast variations have remained undetected in monitoring programmes, or else the sources have a complicated jet structure that is not amenable to our simple analysis. Almost all the sources are seen in a small viewing angle of less than 20 degrees. Our results follow the predictions of basic unification schemes for AGN.}
{}

\authorrunning{Hovatta et al.}
\titlerunning{Doppler factors, Lorentz factors and viewing angles for a sample of AGN}
\maketitle
\section{Introduction}
All radio-bright active galactic nuclei (AGN) have relativistic jets emitting 
synchrotron radiation. 
The jets can at the simplest level be modelled
by using two intrinsic parameters, the Lorentz factor ($\Gamma$), which 
describes the speed of the jet flow, and the viewing angle ($\theta$),
which is the angle between the jet axis and the line of sight to the observer.
These parameters can be calculated if the Doppler boosting factor ($D$) and the
apparent speed $\beta_{\mathrm{app}} = v/c$ are known. We can find out 
the $\beta_{\mathrm{app}}$ from Very Long Baseline Interferometry (VLBI) 
observations, and in the past few 
years major progress has been made in this area \citep[e.g.][]{jorstad01, homan01, kellermann04, jorstad05, piner07, britzen08}. 
The Doppler boosting factors can be 
calculated in various ways, and different methods are compared 
in \cite{lahteenmaki99b} (hereafter LV99). 

A common way to calculate the Doppler boosting factors is to combine
X-ray observations with VLBI component fluxes 
\citep[e.g.][]{ghisellini93, guijosa96, guerra97, britzen07}. 
This method assumes 
inverse Compton (IC) origin of the X-ray emission, and that the same 
synchrotron photons forming the lower frequency radiation are 
responsible also for the IC emission. Assuming that the VLBI observations
are done at the spectral turnover frequency, a predicted 
X-ray flux can be calculated. By comparing this to the observed 
X-ray flux, and by interpreting the excess flux as due to Doppler boosting, 
the Doppler boosting factors can be calculated. 
If the VLBI frequency is not at the turnover, large errors are 
induced in the Doppler boosting factors. This method also suffers 
greatly from non-simultaneous X-ray and VLBI data, 
and as was argued in LV99, gives much less accurate estimates for the 
Doppler boosting factors.  

Using VLBI it is possible to directly observe the brightness temperature 
of the source ($T_{\mathrm{b,obs}}$). This can be compared to the 
intrinsic brightness temperature of the source ($T_{\mathrm{b,int}}$), 
which is often assumed to be the equipartition temperature ($T_{\mathrm{eq}}$) 
\citep{readhead94, lahteenmaki99}. The excess of $T_{\mathrm{b,obs}}$ is 
interpreted as caused by Doppler boosting. This method also 
requires the values to be obtained at the turnover frequency, which 
enhances the errors in the Doppler boosting factors (LV99).

Another way to use VLBI observations is shown in \citet{jorstad05} who 
estimated the variability Doppler boosting factors 
of 15 AGN using Very Long Baseline Array (VLBA) data at 43\,GHz. 
They calculated the flux decline 
time ($\tau_{obs} \propto \tau_{int}D$) of a component in the jet 
and compared it to the measured size of the VLBI component (which does 
not depend on $D$). 
Assuming the intrinsic variability timescale corresponds to the light-travel 
time across the knot, they estimated the Doppler boosting factors. 
They also estimated the Lorentz 
factors and the viewing angles for these sources by using apparent speed data.

Variability timescales can also be obtained from total flux density (TFD)
observations. This is the method used in LV99 and we use the same method 
in our analyses. We decompose each flux curve into 
exponential flares and calculate the variability timescale of each flare. 
From this we gain the observed brightness temperature, which is boosted by 
$D^3$ in comparison with $T_{\mathrm{b,int}}$.
This makes 
possible the extraction of the variability Doppler factor $D_\mathrm{var}$ if 
$T_{\mathrm{b,int}}$ is known. 
In \citet{lahteenmaki99} it was argued, based on observations, 
that in every large
flare $T_{\mathrm{b,int}}$
reaches the equipartition temperature $T_{\mathrm{eq}} = 5\times 10^{10}K$.

In LV99 a sample of 81 sources was studied at 22 and 37\,GHz frequencies. 
They calculated the Doppler boosting factors based on observations from 
a period of over 15 years. For each source they combined the results of the 
two frequency bands and 
chose the fastest flare for the analysis. In 
addition they determined the Lorentz factors and the viewing angles for 
45 sources. We have done similar calculations for a larger sample of sources 
and using data from almost 30 years of monitoring. 
We will 
also compare our results with the results of LV99 to see if the values have 
changed during the past 10 years. 

The paper is organised as follows: in Sect. \ref{sec:data} we describe the 
source sample and the method used. In Sect. \ref{sec:doppler} we calculate 
the variability Doppler boosting factors and compare our results with 
LV99 and other related studies. The Lorentz factors and the viewing angles 
are presented in Sect. \ref{sec:lorentz} and the discussion follows in 
Sect. \ref{sec:discussion}. Finally conclusions are drawn 
in Sect. \ref{sec:conclusions}.

\section{Data and the method}\label{sec:data}
Our sample consists of 87 bright, well-monitored AGN from the Mets\"ahovi
Radio Observatory monitoring list. This is a rather good approximation of a 
complete flux limited sample of 
the brightest compact northern sources. (Missing are a handful of sources for which we had insufficient data to calculate the Doppler factors.) 
The sources have been observed regularly for almost 30 years 
at 22 and 37\,GHz with the Mets\"ahovi 14m telescope \citep{salonen87, terasranta92, terasranta98, terasranta04, terasranta05}. 
Details of the observation method 
and data reduction process are described in \cite{terasranta98}.
Our study also includes unpublished data at 37\,GHz from December 2001 until 
the end of 2006. Data of BL Lacertae objects at 37\,GHz from 
December 2001 until April 2005 are published in \citet{nieppola07}.
In our sample we have 30 high polarisation quasars (HPQs), which have 
optical polarisation exceeding 3 percent at some point in the past.
In addition we have 22 low polarisation quasars (LPQs), 8 quasars (QSOs) for 
which no polarisation data were available, 
22 BL Lacertae objects (BLOs), and 5
radio galaxies (GALs). 
We have used the classification used in the MOJAVE\footnote{http://www.physics.purdue.edu/MOJAVE/} programme whenever possible. 
All the quasars in our sample can be classified as flat spectrum radio 
quasars (FSRQs), which show blazar-like properties.
Our sample does not contain ordinary quasars that have larger viewing 
angles and which do not show the rapid variability common for FSRQs.
In our analyses we have usually combined QSOs with LPQs, 
following the choice of LV99.

\begin{figure}
\resizebox{\hsize}{!}{\includegraphics{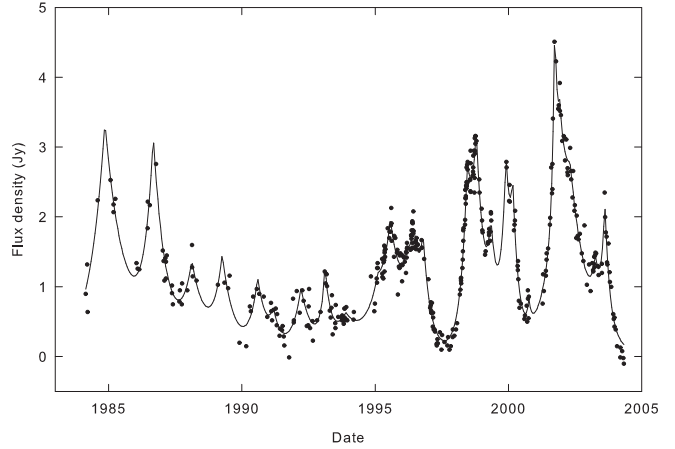}}
\caption{Flux curve of the HPQ source \object{1156+295} (points) decomposed into exponential flares (solid line) at 22\,GHz}
\label{1156_example}
\end{figure}

We decomposed the flux curves into exponential flares of the form
\begin{equation}\label{eq:form}
\Delta S(t) = \left\{ \begin{array}{ll} 
\Delta S_{\mathrm{max}}e^{(t-t_{\mathrm{max}})/\tau}, & t<t_{\mathrm{max}},\\ 
\Delta S_{\mathrm{max}}e^{(t-t_{\mathrm{max}})/1.3\tau}, & t>t_{\mathrm{max}},
\end{array} \right.
\end{equation}
where $\Delta S_{\mathrm{max}}$ is the maximum amplitude of the flare in 
janskys, $t_{\mathrm{max}}$ is the epoch of the flare maximum and 
$\tau$ is the rise time of the flare.
The method is described in detail in \cite{valtaoja99}. 
An example of a fit is presented 
in Fig. \ref{1156_example}, where the flux curve of a HPQ source 
\object{1156+295} is decomposed into exponential flares. The solid line 
represents the fitted sum curve of the individual components and the points 
are actual observed data. We can see that the overall correspondence 
between the two is very good. In LV99 and \cite{savolainen02} it was 
also shown that the individual exponential flare components
correspond very well to the emergence of new VLBI components,
indicating that these flares obtained from the fits
are indeed related to the actual jet physics.

From the fits we obtain the 
necessary parameters to calculate the observed variability brightness
temperature of the source $T_{\mathrm{b,var}}$ (in the source proper frame) 
\begin{equation} \label{eq:T_b}
 T_{\mathrm{b,var}} = 1.548 \times 10^{-32}\frac{\Delta S_{\mathrm{max}}d_\mathrm{L}^2}{\nu^2\tau^2(1+z)},
\end{equation} 
where $\nu$ is the observed frequency in GHz, $z$ is the redshift, $d_\mathrm{L}$ is the luminosity distance in metres, and
$\Delta S_{\mathrm{max}}$ and $\tau$ are defined in Eq. \ref{eq:form}. 
The numerical factor in 
Eq. \ref{eq:T_b} corresponds to using $H_0 = 72 \,\mathrm{km}\,\mathrm{s}^{-1}\,\mathrm{Mpc}^{-1}$, $\Omega_m = 0.27$ and $\Omega_\Lambda = 0.73$, and to assuming that the source is a homogeneous sphere. In the calculation of the 
luminosity distances we have made use of the python version of the 
cosmology calculator created by Edward L. Wright\footnote{http://www.astro.ucla.edu/~wright/CosmoCalc.html} \citep{wright06}. 

The variability Doppler factor can then be calculated as
\begin{equation} \label{eq:D}
D_{\mathrm{var}} = \left[\frac{T_{\mathrm{b,var}}}{T_{\mathrm{b,int}}}\right]^{1/3}.
\end{equation}
We used the value $T_{\mathrm{b,int}} = 5 \times 10^{10}\mbox{K}$. This
value was suggested by \cite{readhead94} and the use of it was 
justified 
in \cite{lahteenmaki99}. Based on simulations, \cite{kellermann04} also 
find the $T_{\mathrm{b,int}}$ to be of the order of $10^{11}\mbox{K}$. 

By combining the Doppler boosting factors with apparent superluminal 
component velocities $\beta_{\mathrm{app}}$, obtained using VLBI, 
we can calculate the variability Lorentz factors 
$\Gamma_{\mathrm{var}}$ and the viewing angles $\theta_{\mathrm{var}}$ 
by using Eqs. \ref{eq:gamma} and \ref{eq:theta}.

\begin{equation} \label{eq:gamma}
\Gamma_{\mathrm{var}} = \frac{\beta_{\mathrm{app}}^2 + D_{\mathrm{var}}^2 + 1}{2D_{\mathrm{var}}}
\end{equation}

\begin{equation} \label{eq:theta}
\theta_{\mathrm{var}} = \arctan\frac{2\beta_{\mathrm{app}}}{\beta_{\mathrm{app}}^2 + D_{\mathrm{var}}^2 - 1}
\end{equation}

We obtained 67 $\beta_{\mathrm{app}}$ values from the MOJAVE sample, 
observed with the VLBA at 15\,GHz. The values are taken from the website on 
September 9, 2008 and some of them may still be preliminary and change 
slightly in the final results (Lister et al. in preparation). 
This is the most homogeneous and largest sample
of $\beta_{\mathrm{app}}$ available at higher radio frequencies.
These values represent the fastest reliable speed in each jet as 
measured by the MOJAVE programme at 15 GHz, using VLBA data spanning 
between 5 to 13 years, depending on the individual source.

\section{The variability Doppler boosting factors}\label{sec:doppler}
\subsection{Estimation of the Doppler boosting factors}\label{sec:estimation}
We were able to calculate the Doppler boosting factor ($D_\mathrm{var}$) 
for 86 sources at 22\,GHz, and for 72 sources at 37\,GHz. 
For many sources we were able to 
determine the $D_\mathrm{var}$ for more than one flare. 
The medians of $D_\mathrm{var}$ are slightly larger at 22\,GHz than at 37\,GHz,
which could be due to different intrinsic brightness temperatures at the two
frequency bands. In our analysis we chose the fastest flare of each source 
(either at 22 or 37\,GHz, whichever had the fastest flare) to calculate 
the $D_\mathrm{var}$, as was done in LV99. The argument for using the fastest 
flare to determine $D_\mathrm{var}$ is that they are most likely to reach 
the limiting brightness temperature and least likely to suffer from 
blending of flares which tends to increase the fitted timescale.
This way we were able to calculate the $D_\mathrm{var}$ 
for all the 87 sources in our sample.
By visual examination, we divided the fits into three categories 
based on the goodness of the fit. No single numerical value, such 
as the $\chi^2$ test, is alone suitable for describing the goodness, 
because these 
usually characterise the entire flux curve, while we have only used one, 
fastest flare, to determine the $D_\mathrm{var}$. In addition, relatively 
large error bars in some fainter sources cause the $\chi^2$ value to be 
small, while we consider a fit to be better when the scatter among the 
datapoints is small.

We classified the $D_\mathrm{var}$ 
as excellent (E, 21 sources), good (G, 24 sources) or 
acceptable (A, 42 sources). In the fits classified as excellent, 
the exponential
decomposition follows the datapoints quite precisely, as in the 
flares after the 
year 1997 in Fig. \ref{1156_example}. The fit is also unambiguous and 
other functions do not describe the behaviour as well. Larger flares in 
Fig. \ref{1156_example} 
before 1995 would mainly be defined as good, because in these flares the 
fit follows the flux curve well, but there is also some scatter, and in some 
cases there are not as many datapoints to define the fit as in the ones 
defined as excellent. In the fits classified as acceptable, the scatter 
around the fit is still larger. 
This is often the case when the flux level of the 
source is modest and the errorbars consequently large. However, none of our 
results change significantly if we exclude the acceptable sources.
All Doppler boosting factors are shown in Table \ref{table:sourcelist}\addtocounter{table}{1}
where the B1950-name, other commonly used name, type of the object, frequency of the $D_\mathrm{var}$ determination, redshift, quality of the $D_\mathrm{var}$, $\log T_{\mathrm{b,var}}$, $D_\mathrm{var}$, $\beta_\mathrm{app}$, $\Gamma_\mathrm{var}$, $\theta_\mathrm{var}$, core dominance parameter $R$, and maximum optical polarisation $P_\mathrm{max}$ and its reference are listed. 

It is difficult to determine exact error estimates 
for the $D_\mathrm{var}$ of each source. Some indications can be obtained 
from the standard deviation of $D_\mathrm{var}$ calculated from the various 
flares in one source. We calculated the deviations for all the sources 
classified as excellent, which had more than one flare to determine 
the $D_\mathrm{var}$, including also flares determined as good. 
There were 45 such cases (including 
all the fits at both 22 and 37\,GHz) and, on average, each source had 5.7 well-defined 
flares. The median standard deviation for these is $\sim 27\%$, which can be 
thought of as an indication of the upper limit for the error estimate since 
in many cases the change 
in the $D_\mathrm{var}$ of individual flares can also be due to 
differences in the 
source behaviour. It is also more likely to see a very fast flare in each 
source the longer they are monitored.
More insight into the 
errors can be obtained when our new 
$D_\mathrm{var}$ are compared to other studies (cf. Chapt. \ref{sec:old}).
 
\begin{table}
\centering
\caption[]{Median values of $\log(T_{\mathrm{b,var}})$ and $D_\mathrm{var}$.}
\label{median_D_T}
\begin{tabular}{lccc}
\hline
\hline
\noalign{\smallskip}
Type 	  & 	 N 	  & 	 $\log(T_{\mathrm{b,var}}) [K]$ 	  & 	 $D_\mathrm{var}$ 	  \\
 \noalign{\smallskip} 
\hline 
\noalign{\smallskip} 
HPQ 	  & 	 30 	  & 	 14.31 	  & 	 15.98 	  \\
 LPQ 	  & 	 30 	  & 	 13.92 	  & 	 11.90 	  \\
 FSRQ       &      60       &      14.19   &      14.61    \\
 BLO 	  & 	 22 	  & 	 13.09 	  & 	 6.25 	  \\
 GAL 	  & 	 5 	  & 	 11.65 	  & 	 2.07 	  \\
 ALL 	  & 	 87 	  & 	 13.94 	  & 	 12.02 	  \\
 \noalign{\smallskip} 
\hline 
\end{tabular}
\end{table}

\begin{figure}
\resizebox{\hsize}{!}{\includegraphics{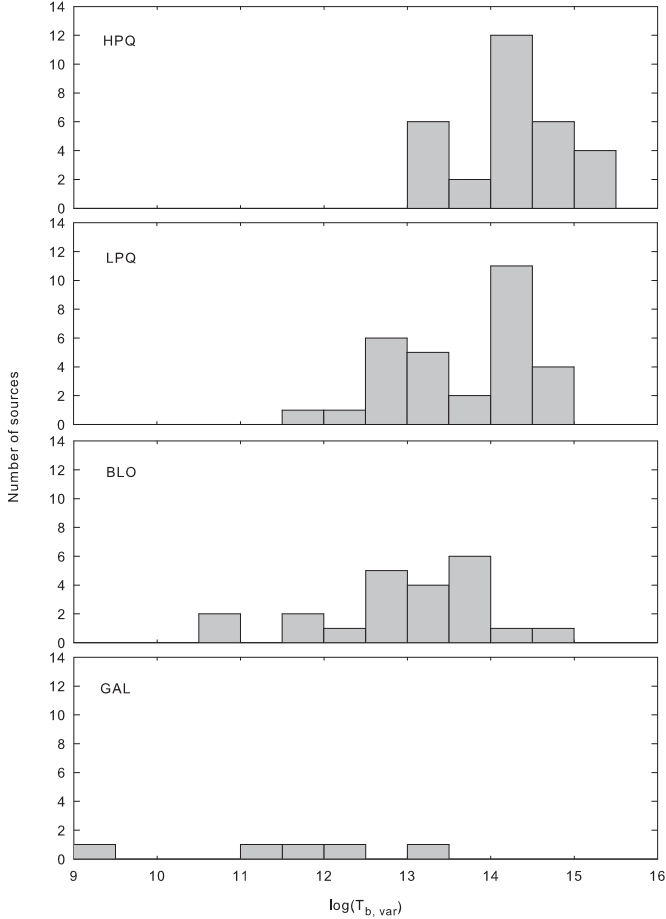}}
\caption{Distribution of $T_{\mathrm{b,var}}$ of the fastest flare in each source}
\label{histo_T}
\end{figure}

\begin{figure}
\resizebox{\hsize}{!}{\includegraphics{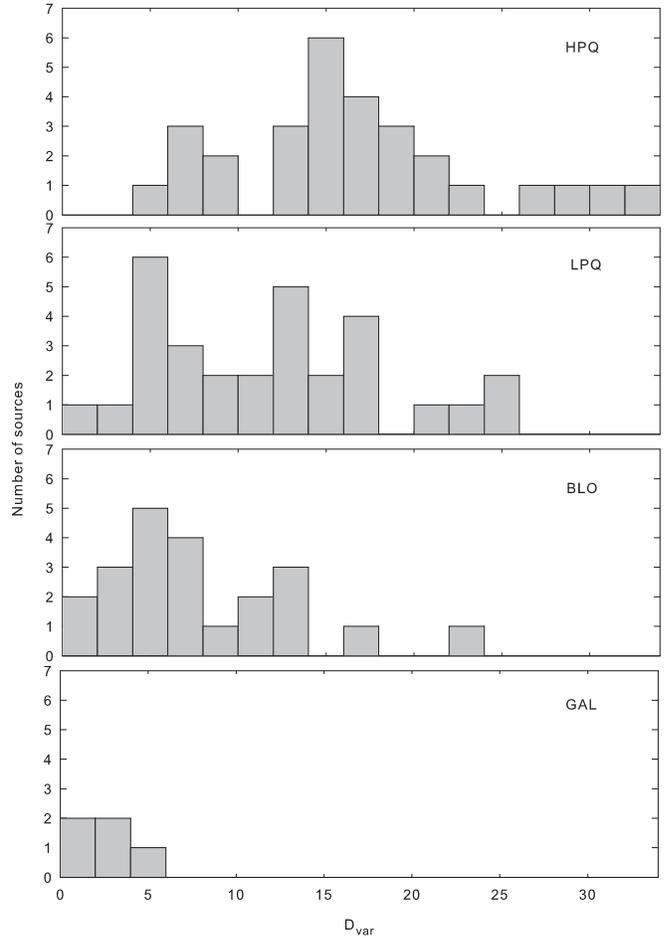}}
\caption{Distribution of $D_\mathrm{var}$ of the fastest flare in each source}
\label{histo_D}
\end{figure}

In Table \ref{median_D_T} we show the median values of $\log(T_{\mathrm{b,var}})$ and 
$D_\mathrm{var}$ for each source class separately and for HPQs and LPQs 
combined together (FSRQ).
 The distributions are 
shown in Figs. \ref{histo_T} and \ref{histo_D} and we can see that the distributions of quasars and BLOs have considerable overlap. HPQs 
have a tail extending to higher $D_\mathrm{var}$ and LPQs seem to be in between  
the HPQs and BLOs. Also, it is interesting to note that all the quasars are
clearly Doppler-boosted, the smallest estimated $T_{\mathrm{b,var}}$ being  
$3.5\times 10^{11}$ K for \object{1928+738}.
We ran the Kruskal-Wallis analysis to examine the differences between 
source classes. (All Kruskal-Wallis analyses in this 
paper have been performed with the Unistat statistical package for 
Windows\footnote{http://www.unistat.com/} (version 5.0).) 
The results confirm that all the source classes differ from the other  
classes significantly with a 95\% confidence limit. 

\subsection{Comparison with previous analyses}\label{sec:old}
\begin{figure}
\resizebox{\hsize}{!}{\includegraphics{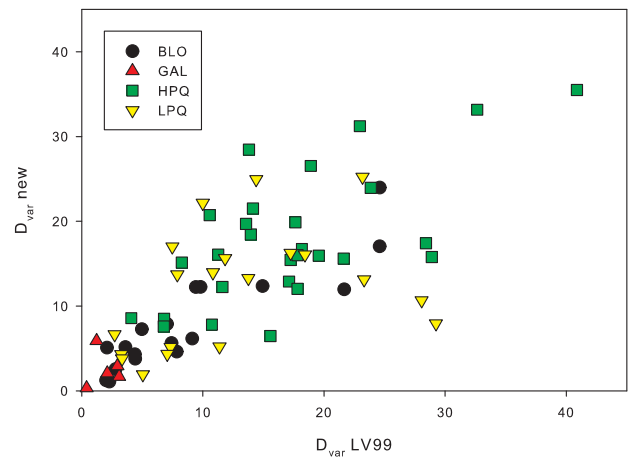}}
\caption{Correlation between the $D_\mathrm{var}$ from LV99 and our new $D_\mathrm{var}$ values}
\label{old_vs_new}
\end{figure}

Our sample has 71 sources in common with the sample of LV99. We have 
re-calculated the Doppler boosting factors of LV99, using the current 
cosmological model. Figure \ref{old_vs_new} shows
the correlation between the Doppler boosting factors of LV99 and the new
values. The results are very similar and the values correlate with
a coefficient r = 0.77 (p=0.0000). Kruskal-Wallis analysis also shows 
that the values come from the same population. This confirms that the 
method is 
reliable because the results have not changed even though 
we have now 
ten more years of data.  The differences 
are mainly due to poor fits in LV99 which are due to poor sampling or low 
flux density (large scatter) in the data. The scatter between the old and the 
new 
values is consistent with the error analysis in Sect. \ref{sec:estimation}.
In some cases (e.g. \object{0430+052} and 
\object{1156+295}) the source has clearly changed its behaviour and exhibits 
a much faster flare in our new dataset.
The new estimates which are calculated using almost 30 years of data 
should therefore be more representative of the source behaviour.

\begin{figure}
\resizebox{\hsize}{!}{\includegraphics{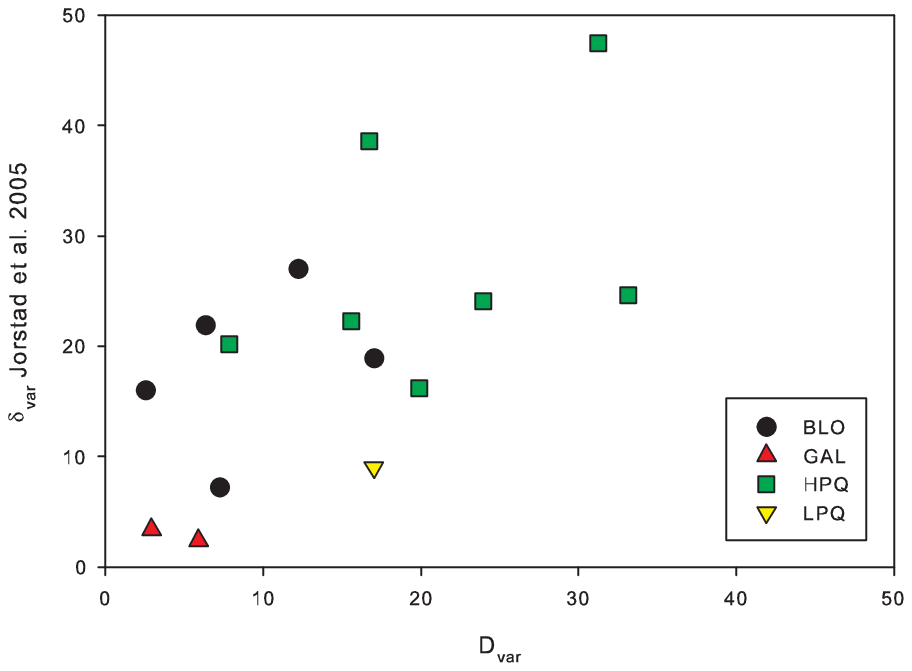}}
\caption{Correlation between the $D_\mathrm{var}$ and $\delta_\mathrm{var}$ from \cite{jorstad05}.}
\label{D_vs_jorstad}
\end{figure}

We also compared our $D_\mathrm{var}$ values with the \cite{jorstad05} 
values for 15 AGN, obtained at 43\,GHz. Figure \ref{D_vs_jorstad} shows the 
correlation between the two values, and a Spearman rank correlation gives 
a coefficient r=0.56 (p=0.0123).
We can see that the values in \cite{jorstad05} are in general somewhat higher 
than ours. This can be due to their higher observing frequency (cf. Discussion). Also,
their analysis method gives only an upper limit to some sources. The 
distribution of source classes is similar to ours, with quasars 
having the highest 
Doppler boosting factors, GALs the lowest and BLOs being in between them.
Therefore
we conclude that the Doppler boosting factors of these two analyses 
correspond well with each other.

\cite{homan06} argued that during the most active state 
$T_{\mathrm{b,int}}$ should be closer to $2 \times 10^{11}\mathrm{K}$ and 
therefore the $D_\mathrm{var}$ of the LV99 are overestimated. 
This would make our $D_\mathrm{var}$ values even smaller, and the 
correspondence to \cite{jorstad05} would be worse. 
Higher $T_{\mathrm{b,int}}$ would also increase our Lorentz factors for the 
fastest sources, and 
as is shown later (cf. Chapters \ref{sec:lorentz} and \ref{sec:discussion}), 
our new values are already twice as high as in LV99 and in some 
sources even extremely high. We also note that a different value for 
$T_{\mathrm{b,int}}$ does not change the distributions themselves, only the 
numerical values.

We also compared our Doppler boosting factors to a recent study at a lower 
frequency of 5\,GHz \citep{britzen07}. They calculated the IC
Doppler boosting factors by using VLBI data from the Caltech-Jodrell Bank 
Flat-Spectrum source sample \citep{taylor96} and non-simultaneous 
ROSAT X-ray data. The 
Spearman rank correlation between our 24 common sources (excluding one outlier, \object{0836+710}, with $D_\mathrm{IC}=88$) is still quite good 
(r=0.63, p=0.0004), and the slope of the linear fit between $D_\mathrm{IC}$ and $D_\mathrm{var}$ is almost exactly one. We believe that most of the scatter 
is due to the errors in the IC Doppler boosting factors, since LV99 showed 
that for several reasons these are likely to be much less accurate than 
the variability Doppler boosting factors. 

\subsection{Core dominance}\label{sec:coredominance}
\begin{figure}
\resizebox{\hsize}{!}{\includegraphics{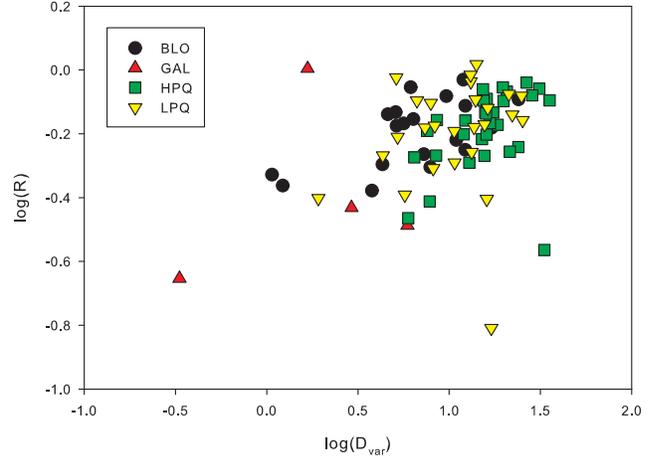}}
\caption{Correlation between $\log(R)$ using data from \cite{kovalev05} and $\log(D_\mathrm{var})$, excluding the outlier source \object{0923+392} with $\log (R) = -1.76$ and $\log (D_\mathrm{var}) = 0.63$.}
\label{cored}
\end{figure}

Standard beaming models expect that more core-dominated objects should 
be more beamed and thus have higher Doppler boosting factors. We  
studied this by calculating the core-dominance parameter R from VLBA 
data of \cite{kovalev05} at 15\,GHz. The core-dominance is calculated 
by relating the flux density of the core $S_{\mathrm{core}}$ to the total 
single-dish flux density observed at 15\,GHz $S_{\mathrm{tot}}$. 
In \cite{kovalev05} 
these are given for 250 sources observed with the VLBA at 15\,GHz at different 
epochs. 
Their sample includes 80 sources for which we have 
determined $D_\mathrm{var}$. We calculated the median R for each 
source from the separate epochs. Figure \ref{cored} shows the 
correlation between $\log\mathrm{R}$ and $\log D_\mathrm{var}$, 
excluding an outlier 
source \object{0923+392} with a very small core dominance ($\log R = -1.76$). 
Spearman 
rank correlation between the parameters is $r=0.37$ $(p=0.0004)$. 
When the outlier source is 
included, the correlation is still significant with a coefficient 
$r=0.39$ $(p=0.0002)$. This shows 
that there is indeed some indication that sources which are more core-dominated
are also more boosted. In \cite{kovalev05} the core-dominance is defined 
to be the relation of $S_{\mathrm{core}}$ to the total VLBA flux 
density $S_{\mathrm{VLBA}}$. 
We tested the correlation using also this parameter but the results did not 
change because $S_{\mathrm{tot}}$ and $S_{\mathrm{VLBA}}$ are so similar. 
Using the core-dominance defined with $S_{\mathrm{VLBA}}$, 
\cite{kovalev05} show that quasars and BLOs are significantly different from 
GALs with lower core-dominance.

Similar calculations were made in \cite{britzen07} 
for their sample. They used the core flux density and total 
single-dish flux density at 
5\,GHz to calculate the core-dominance parameter. 
They found no significant correlation between their 
core-dominance parameter and the IC Doppler boosting factor. We have only 25
sources in common with their sample, and when we compared our $D_\mathrm{var}$
with their core-dominance parameter, we found no correlation.

\section{The Lorentz factors and viewing angles}\label{sec:lorentz}
\begin{figure}
\resizebox{\hsize}{!}{\includegraphics{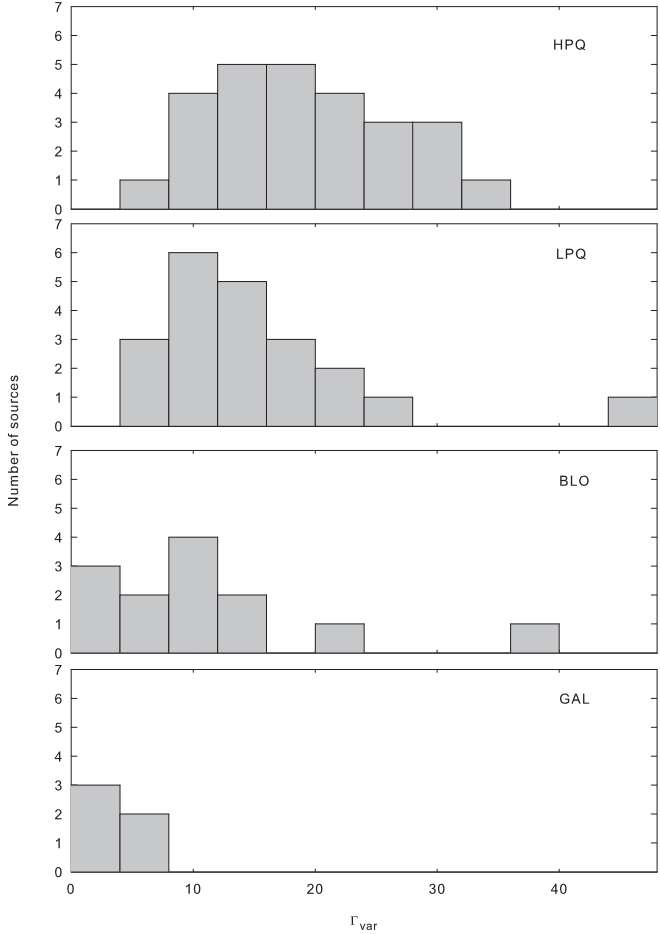}}
\caption{Distribution of $\Gamma_{\mathrm{var}}$ of the fastest flare in each source (excluding the outlier LPQs 0923+392 and 1730-130)}
\label{hist_gamma}
\end{figure}

\begin{figure}
\resizebox{\hsize}{!}{\includegraphics{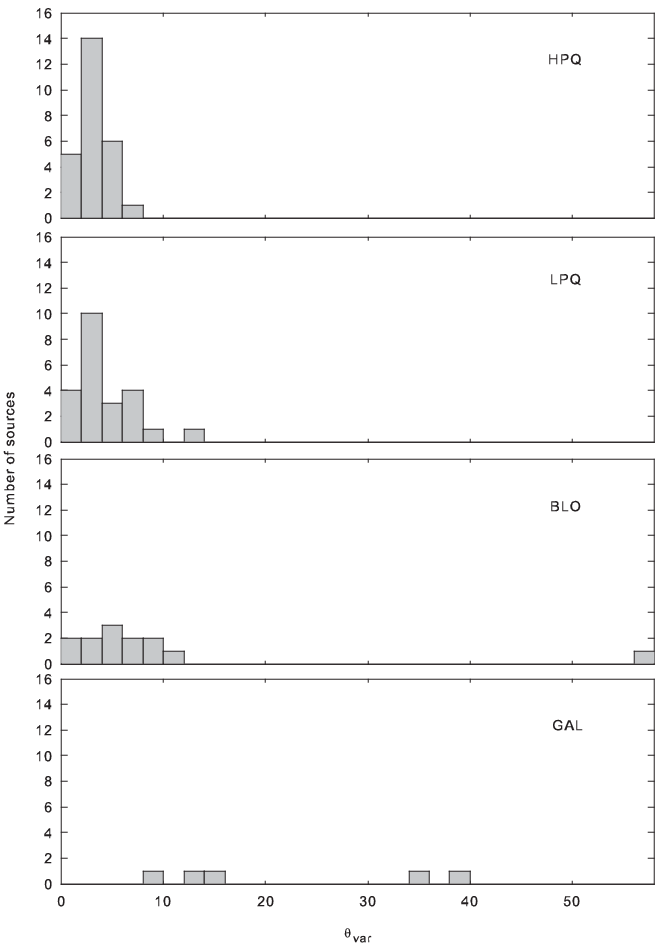}}
\caption{Distribution of $\theta_{\mathrm{var}}$ of the fastest flare in each source}
\label{hist_theta}
\end{figure}

We were able to calculate Lorentz factors $\Gamma_{\mathrm{var}}$ and 
viewing angles $\theta_{\mathrm{var}}$ for 67 sources. The median values 
of different source classes are shown in Table \ref{median_G_Th}. These are 
affected by two outliers with exceptionally large $\Gamma_{\mathrm{var}}$. 
The source \object{0923+392} has $\Gamma_{\mathrm{var}} = 216.1$ and 
\object{1730-130} has $\Gamma_{\mathrm{var}} = 64.6$. Both of these 
show high superluminal motion of $\beta_{\mathrm{app}} > 35c$, which increases 
the Lorentz factors. The distributions of the source classes 
are shown in Figs. \ref{hist_gamma} and 
\ref{hist_theta}. In $\Gamma_{\mathrm{var}}$ the distributions of quasars and 
BLOs overlap, but BLOs and GALs have slower jet speeds than quasars. 
Kruskal-Wallis analysis shows that when $\Gamma_{\mathrm{var}}$ are 
studied without the outlier sources, HPQs differ from other classes with 
higher Lorentz factors, and GALs differ with smaller Lorentz factors. BLOs and LPQs come from the same population with a 13\% confidence.
There is one BLO (\object{1823+568}) for which $\Gamma_{\mathrm{var}} = 37.8$.
This source has also been classified as a HPQ \citep[e.g.][]{veron06} 
and therefore we ran the 
KW-analysis again by moving this source into the HPQ class. In this case also 
BLOs and LPQs differ significantly from each other with a 96\% confidence. 
We must note that in our samples of 13 to 26 objects the significance of 
differences can depend on the classification of a single extreme source.
However, a clear result is that the BLOs and the quasars differ from 
each other with BLOs having slower jets (FSRQs and BLOs differ significantly with a 99\% confidence if \object{1823+568} is classified as a BLO and with a 99.9\% confidence if \object{1823+568} is classified as a HPQ). 
This result is in accordance with 
several earlier but more indirect estimates of jet speeds. 
Similar results have also been obtained 
with simulations \citep{hughes02}.
 
When $\theta_{\mathrm{var}}$ is studied, the distributions overlap even more 
and KW-analysis shows that GALs and BLOs differ from other source classes 
with a 95\% confidence. Also, if the differences between FSRQs and BLOs are 
studied, they differ from each other significantly with a 99\% confidence with 
BLOs having larger viewing angles. 

\begin{table}
\centering
\caption[]{Median values of $\Gamma_{\mathrm{var}}$ and $\theta_{\mathrm{var}}$.}
\label{median_G_Th}
\begin{tabular}{lrrr}
\hline
\hline
\noalign{\smallskip}
Type 	  & 	 N 	  & 	 $\Gamma_\mathrm{var}$ 	  & 	 $\theta_\mathrm{var}$ 	  \\
 \noalign{\smallskip} 
\hline 
\noalign{\smallskip} 
HPQ 	  & 	 26 	  & 	 17.41 	  & 	 3.28 	  \\
 LPQ 	  & 	 23 	  & 	 13.96 	  & 	 3.90 	  \\
  	  & 	 $21^a$ 	  & 	12.65  	  & 	3.96  	  \\
FSRQ       &      49       &      16.24    &      3.37     \\
 BLO 	  & 	 13 	  & 	 10.29 	  & 	 5.24 	  \\
 GAL 	  & 	 5 	  & 	 1.82 	  & 	 15.52 	  \\
 ALL 	  & 	 67 	  & 	 13.96 	  & 	 3.81 	  \\
 \noalign{\smallskip} 
\hline 
\end{tabular}
\begin{list}{}{\setlength{\leftmargin}{45pt}}
\item \footnotesize{$^a$ = excluding outliers 0923+392 and 1730-130}
\end{list}
\end{table}

\begin{figure}
\resizebox{\hsize}{!}{\includegraphics{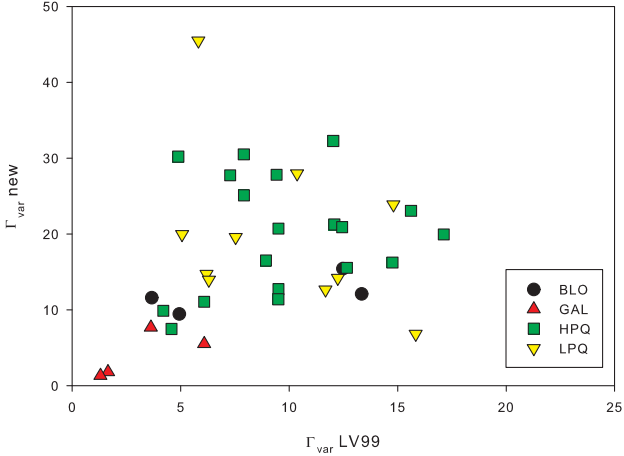}}
\caption{Correlation between Lorentz factors from LV99 and the new values (excluding the outlier source 0923+392 with a Lorentz factor of 216 in the new analysis).}
\label{gamma_new_old}
\end{figure}

We compared our new Lorentz factors with the ones from LV99. There are 
38 sources in common in our samples, and in Fig. 
\ref{gamma_new_old} we can see that the new $\Gamma_\mathrm{var}$ are about 
twice as large as in LV99. This is mainly because also the apparent speeds
used in our analysis are about twice as large as in LV99, where they were 
collected from the literature. We should now have better and more homogeneous 
estimates of $\beta_{\mathrm{app}}$ which should make our new estimates 
more accurate. The correlation between our $\Gamma_\mathrm{var}$ and those 
from LV99 is only $r=0.37$ $(p=0.0113)$ when the outlier (\object{0923+392}) 
is not included (the other outlier \object{1730-130} is not included in the 
sample of LV99). 
When we compare our $\Gamma_\mathrm{var}$ with Lorentz factors 
estimated in \cite{jorstad05} we find a good correlation of 
$r=0.50$ $(p=0.0281)$ which gives credibility to our new estimates. 
Also the distribution is similar with quasars having the highest 
Lorentz factors, BLOs being in between and GALs having the smallest 
Lorentz factors.
We have only 12 sources in common with the sample of \cite{britzen07} and we 
find no correlation between our Lorentz factors 
even if we leave out their outlier 
source \object{0016+731} for which they determine a Lorentz factor of 860.
The differences are probably due to their much lower values of 
$\beta_{\mathrm{app}}$ and differences in the Doppler boosting factors. 

When studying the viewing angles, the difference between our new and LV99 
values is not as large and the correlation is good ($r=0.51$ $p=0.0005$). 
The correlation between our values and 
\cite{jorstad05} values is also very good with $r=0.59$ $(p=0.0105)$. 
Again there are no significant correlations between $\theta_{\mathrm{var}}$
and values from \cite{britzen07}.

\section{Discussion}\label{sec:discussion}
Although the Doppler boosting factors have remained on the average almost 
identical (cf. Fig. \ref{old_vs_new}) even 
though we now have ten more years of data, there is a factor of two difference
in the Lorentz factors when comparing our new results to LV99. This is mainly 
due to 
twice as fast apparent speeds in our new analysis. In LV99 the 
$\beta_\mathrm{app}$ values are mainly from \citet{vermeulen94}, in which 
the values were gathered from the literature. This means that the speeds were 
obtained in the 1980s and early 1990s from a variety of observing programmes and frequencies, mainly at 
5\,GHz. We use 15\,GHz MOJAVE data from 1994 up to September 2008, 
giving a far more uniform dataset. 
We also note that higher 
frequency VLBI observations often tend to give higher apparent speeds as 
can be seen by comparing, e.g., data from 
\citet{vermeulen94, kellermann04, jorstad05} and \citet{britzen07}, although 
the reason for this is not clear. The use 
of a single frequency therefore diminishes the internal scatter.

Long-term observations, both TFD monitoring 
and VLBI, are essential in understanding source behaviour. We have 
already shown this in our earlier studies of long-term variability behaviour 
of these sources \citep{hovatta07, hovatta08b, hovatta08}. Some sources have 
changed their behaviour in our TFD observations during the past 10 years. 
Similar changes can be seen in VLBI sources over a long time, such as the 
position swings seen, for example, in \object{3C 273} \citep{savolainen06b} 
and \object{NRAO 150} \citep{agudo07}. We emphasise that our way to 
estimate the Lorentz factors and viewing angles is as good as can be, when 
we are characterising a complex, changing jet with just two parameters, 
both assumed to be constant.

In addition to having twice as high apparent speeds in general, there are 
also some extreme sources showing very high apparent motion. While in 
LV99 the highest apparent speed was 14.9, we now have two sources 
(\object{0923+392} and \object{1730-130}) with $\beta_\mathrm{app} > 35c$.
As a consequence, the Lorentz factors of these sources are also extremely 
high, $\Gamma_\mathrm{var} = 216$ for \object{0923+392} and 
$\Gamma_\mathrm{var} = 65$ for \object{1730-130}. We also note that if 
the $\beta_\mathrm{app} = 45.9$ for the source \object{1510-089} from 
\citet{jorstad05} is accepted, our $\log(T_\mathrm{b,var}) = 14.4$ would 
indicate a $\Gamma_\mathrm{var} = 71$. Thus, the existence of a class of 
very fast jets should perhaps at least be considered as a possibility.

However, at least the $\Gamma_\mathrm{var} = 216$ seems
rather unlikely in view of our current knowledge of the jets in AGN. 
One alternative explanation is that \object{0923+392} 
has a higher observed brightness temperature than what we have obtained 
($\log(T_\mathrm{b,var}) = 12.6$) from our monitoring. If we saw changes 
of about 1 Jy within a time period of a week, the brightness temperature 
would be of the order of $10^{15} K$, which would change the Lorentz factor 
to a more acceptable value of under 50. 
Our sampling, like in other monitoring programmes, 
is too sparse to detect such rapid flares reliably, and therefore we have 
initiated a denser monitoring schedule for the source. 
Another possibility is that the source has a complicated internal structure or 
geometry, such that the TFD variations and the apparent speeds do not refer 
to the same component \citep[e.g.][]{alberdi93, fey97, alberdi00}.
However, independent of our 
$T_\mathrm{b,var}$ 
estimates, \object{0923+392} must have a Lorentz factor of at least 43 
as $\Gamma \ge \beta_\mathrm{app}$.

\begin{figure}
\resizebox{\hsize}{!}{\includegraphics{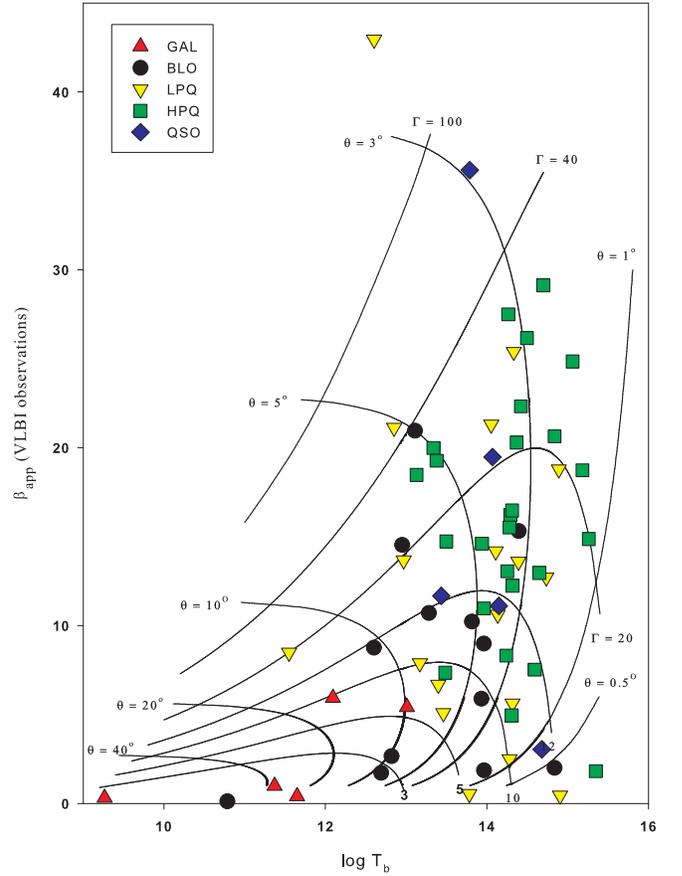}}
\caption{Observable quantities $\beta_\mathrm{app}$ and 
$\log(T_\mathrm{b,var})$ together with intrinsic parameters $\Gamma_\mathrm{var}$ and $\theta_{\mathrm{var}}$.}
\label{Tb_beta}
\end{figure} 

In Fig. \ref{Tb_beta} we have plotted the observable quantities 
$\beta_\mathrm{app}$ and $\log(T_\mathrm{b,var})$. We have 
included curves to mark areas of different $\Gamma_\mathrm{var}$ and 
$\theta_{\mathrm{var}}$. The outlier sources are clearly visible in this 
plot but otherwise the sources are within rather well-defined limits. Almost all the 
sources have $\Gamma_\mathrm{var} < 40$ and $\theta_{\mathrm{var}} < 20^\circ$.
The differences between the source classes are also seen in this plot, with 
GALs having slow speeds and low brightness temperatures. 
Using Monte-Carlo simulations, \citet{cohen07} find an upper limit 
for the Lorentz factor to be $\Gamma \approx 32$. This agrees quite well 
with our results, although in our sample we have five sources with 
$\Gamma_\mathrm{var}$ between 30 and 50 and two quasars with 
$\Gamma_\mathrm{var} > 50$. 

\begin{figure}
\resizebox{\hsize}{!}{\includegraphics{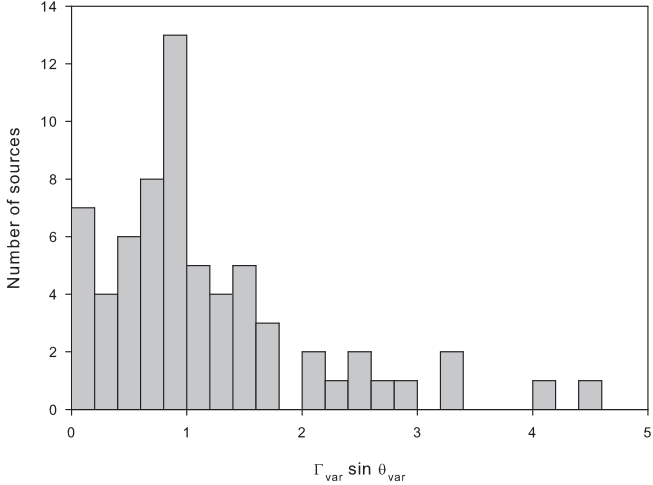}}
\caption{Distribution of $\Gamma \sin\theta$, excluding the source \object{0923+392} with $\Gamma_\mathrm{var} \sin\theta_\mathrm{var} = 10$.}
\label{sintheta}
\end{figure} 

A common assumption is that sources are viewed close to the critical angle 
$\theta_c = 1/\Gamma$. Figure \ref{sintheta} shows the
$\Gamma \sin\theta$ distribution for our sources, which indeed does peak around
1. However, a number of sources have $\Gamma \sin\theta$  significantly 
larger or smaller, so that the assumption  of $\Gamma \sin\theta = 1$  will 
in many cases lead to false conclusions both for individual sources and 
for samples. \citet{cohen07} have calculated a Monte Carlo simulation for 
a flux density limited survey (similar to ours), 
(Fig. 1c in \citet{cohen07}, in which the sources with $\Gamma = 15$ are plotted), finding a distribution rather similar to our empirical one.

\begin{figure}
\resizebox{\hsize}{!}{\includegraphics{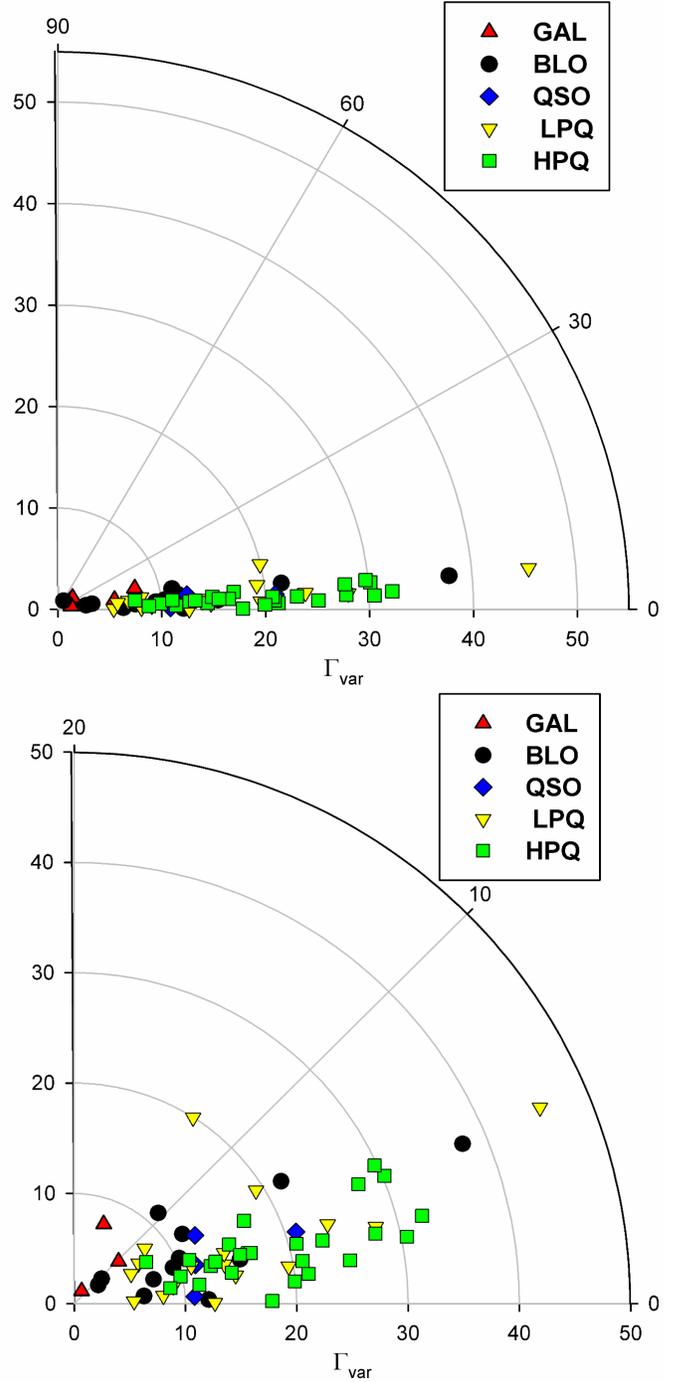}}
\caption{Polar plot of Lorentz factors against the viewing angles, excluding the outlier sources 0923+392 and 1730-130. The lower panel shows only the 0$^\circ$ to 20$^\circ$ portion of the viewing angles, additionally excluding the galaxies 0007+106 and 0316+413, and the BLO 1807+698.}
\label{polar}
\end{figure}

\begin{figure}
\resizebox{\hsize}{!}{\includegraphics{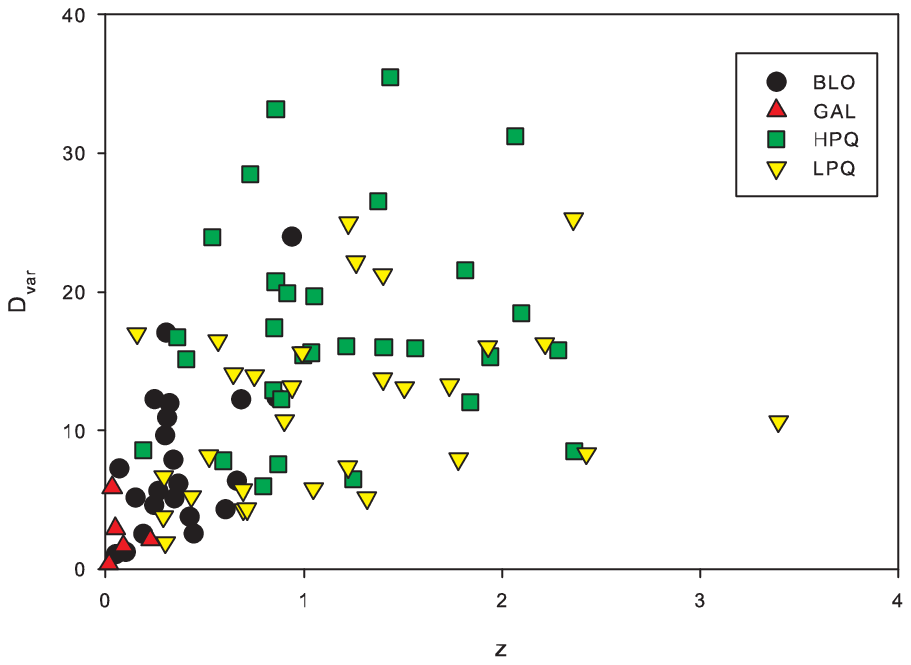}}
\caption{$D_\mathrm{var}$ against the redshift.}
\label{D_z}
\end{figure}

The simplest unification scheme for AGN \citep[e.g.][]{barthel89, urry95} 
predicts that in all radio quasars (including ordinary quasars and FSRQs) 
the $\Gamma$ distributions should be similar. Ordinary quasars, on the other 
hand, should have larger viewing angles than blazars (FSRQs and BLOs). 
BLOs could also have a different $\Gamma$ 
distribution because their parent population is different than in 
quasars. In Fig. \ref{polar} we show the Lorentz factors and viewing 
angles in a polar plot. It is easy to see that almost all the sources are seen 
at a small viewing angle. What must be kept in mind is that our sample 
essentially includes the $\sim 100$ brightest northern compact radio 
sources, so this is as expected for Doppler boosting dominated sources. 
Although we have divided our quasars into HPQs and LPQs, 
they all are FSRQs (or blazars), and we do not see ordinary 
quasars with $\theta_\mathrm{var} > 20^\circ$. Similarly, all our 
BLOs are radio selected BLOs (RBLs) which should have small 
viewing angles \citep{barthel89}. 
In addition, the five GALs in our sample are the most compact and variable 
sources of their type. Figure \ref{D_z} also shows how the sources at 
different redshifts are selected due to their Doppler boosting so that 
small Doppler boosting factors are not seen at high redshifts. These 
caveats must be kept in mind when comparing our different classes of 
sources with each other.

The distribution of $\Gamma_\mathrm{var}$ agrees well with unification 
schemes for AGN. \cite{padovani92} and \cite{urry91} calculated beamed luminosity functions for FSRQs and BLOs, and compared them with 
observed luminosity functions, which were assumed to have the same 
shape as the intrinsic luminosity function of radio-loud quasars. The validity 
of this assumption was recently confirmed by using the maximum likelihood 
method to calculate the intrinsic luminosity functions \citep{liu07, cara08}. 
In \citet{padovani92} the FSRQs were best described with a 
distribution of $5 \lesssim \Gamma \lesssim 40$ with an average of $\sim 11$
(using a cosmology with $H_0 = 50 \,\mathrm{km}\,\mathrm{s}^{-1}\,\mathrm{Mpc}^{-1}$ and $q_0 = 0$).
In our sample all the FSRQs have  
$\Gamma_\mathrm{var} > 5$ and three have $\Gamma_\mathrm{var} > 40$. 
The median 
$\Gamma_\mathrm{var}$ is somewhat bigger than the average from the luminosity 
functions. In \cite{urry91} the BLOs were best described with a distribution 
of $5 \lesssim \Gamma \lesssim 35$ with an average of $\sim 7$. Again we 
have only three BLOs with $\Gamma_\mathrm{var} < 5$ and one BLO (\object{1823+568}, also classified as HPQ) with $\Gamma_\mathrm{var} > 35$. Our median 
is also only 
slightly bigger. In addition they made another fit for the BLOs with a 
distribution of $2 \lesssim \Gamma \lesssim 20$ which also fit the data 
well. In this case only one BLO (\object{1807+698} with $\Gamma_\mathrm{var} = 1.0$) would be below the lower limit and
there would be two BLOs with $\Gamma_\mathrm{var} > 20$.
Their calculations also set a value for the critical 
viewing angle in which all the FSRQs and BLOs should be seen. For 
FSRQs the angle is $14^\circ$, and all our quasars are within this limit. 
For BLOs the critical angle is either $11^\circ$ or $19^\circ$ depending 
on the model. In both cases only one BLO (\object{1807+698} with 
$\theta_{\mathrm{var}} = 57^\circ$, which is also the nearest BLO in our 
sample) would 
have a viewing angle outside the limit. 

\begin{figure}
\resizebox{\hsize}{!}{\includegraphics{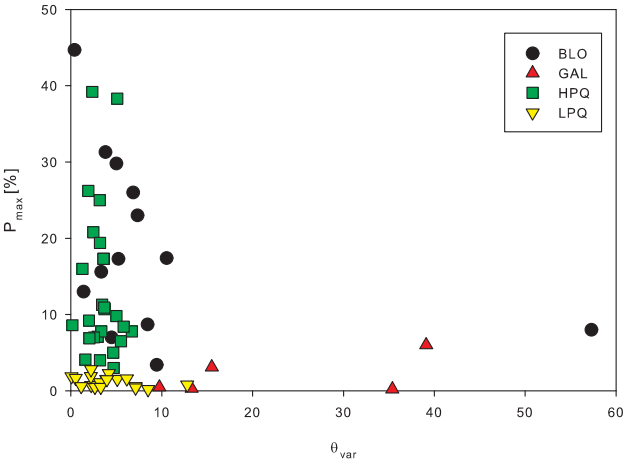}}
\caption{Maximum optical polarisation against the viewing angle.}
\label{Pmax}
\end{figure} 

As expected with sources with small viewing angles, distributions of 
LPQs and HPQs overlap and one should not identify LPQs as ordinary quasars.
In Fig. \ref{Pmax} we show the maximum optical polarisation against 
the viewing angle for all the sources for which we could find polarisation 
information in the literature ($P_\mathrm{max}$ and its reference for 
all the sources are shown in Table \ref{table:sourcelist}). 
Most of the sources 
seem to be within an envelope with $P_\mathrm{max}$ decreasing as 
$\theta_\mathrm{var}$ increases. The situation may be similar as in 
Seyfert galaxies with an obscuring torus \citep[e.g.][]{schmitt01}. 
When the viewing angle is small, we can see deeper into the 
jet and are more likely to see the source as more optically polarised. 
 Also, it has been shown that both the optical and the radio 
polarisation originate from transverse shocks located very close to the base of 
the jet \citep{lister00}.
Therefore the 
range of polarisation variability is much higher for sources with a 
viewing angle of a few degrees than for sources with $\theta_\mathrm{var} = 
10^\circ - 20^\circ$.
In Fig. \ref{Pmax} it is 
also possible to see that the definition $P_\mathrm{max} = 3\%$ for an 
object to be classified as highly polarised is somewhat artificial. We 
believe that with a larger number of polarisation observations, a considerable 
fraction of the LPQ population could, at times, show polarisation 
exceeding 3\%, as expected from blazar-type AGN.
Since there usually is only 
a small number of polarisation observations, the likelihood of observing 
high polarisation increases with decreasing viewing angle. 
However, as has been pointed out by \citet{lister00}, 
there is a physical difference between at least some HPQs and 
LPQs, especially in the magnetic field structure of their inner jets. 

Figure \ref{Pmax} hints that the transition from blazar-type AGN to 
ordinary quasars may occur around a viewing angle of 15$^\circ$ to 20$^\circ$, 
presumably corresponding to the half opening angle of an AGN obscuring 
torus, and in agreement with, e.g., estimates from luminosity functions and 
source counts. As our sample does not contain any ordinary quasars, no 
definite conclusions can be drawn.

\section{Conclusions}\label{sec:conclusions}
We have decomposed flux curves of 87 sources into exponential flares 
and using the fits calculated the variability brightness temperature and 
Doppler boosting factor for each source. In addition we used new MOJAVE 
observations of apparent jet speeds to calculate the variability 
Lorentz factors and viewing angles. We have compared our results with LV99, 
in which the parameters were determined in a similar 
way. In our new analyses we have used almost 15 years more data and the 
estimates should be more accurate. Our main conclusions can be summarised 
as follows.
\begin{enumerate}
\item The variability Doppler boosting factors have remained on the average almost identical compared to LV99. All the quasars are Doppler-boosted and they are in general more boosted than the BLOs or GALs.

\item The Lorentz factors of our new analyses are about twice as large as in LV99. The difference can be explained with twice as large apparent VLBI speeds in our new analyses. The BLOs have slower jets compared to quasars ($\Gamma_\mathrm{BLO} = 10.3$, $\Gamma_\mathrm{FSRQ} = 16.2$ and $\Gamma_\mathrm{GAL} = 1.8$).

\item A few sources, \object{0923+392} and \object{1730-130}, have extreme Lorentz factors (216 and 65, respectively). Either these Lorentz factors are real, or the sources exhibit so rapid flares that the fast variations have remained undetected in the monitoring programmes. We are studying the second possibility with dense observations of \object{0923+392}. A third possibility is that the sources have a structure too complex for our method.

\item Almost all the sources in our sample are seen in a small viewing 
angle of less than 20 degrees ($\theta_\mathrm{BLO} = 5.2^\circ$, $\theta_\mathrm{FSRQ} = 3.4^\circ$ and $\theta_\mathrm{GAL} = 15.5^\circ$).

\item The viewing angle distribution peaks around $\Gamma \sin\theta = 1$, with a distribution similar to that found in simulations.

\item Our results generally follow the predictions of basic unification models for AGN. Based on our results, we cannot separate HPQs and LPQs from 
each other, and therefore it is well-grounded to treat them as a single 
group of FSRQs when the jet parameters are considered.

\end{enumerate}

\begin{acknowledgements}
We thank M. Lister and the members of the MOJAVE team for providing data in
advance of publication. We acknowledge the support of the Academy of Finland (project numbers 212656 and 
210338). This research has made use of data from the MOJAVE (Lister and Homan, 2005, AJ, 130, 1389) and 2cm Survey (Kellermann et al., 2004, ApJ, 609, 539) programs.

\end{acknowledgements}

\bibliographystyle{/home/tho/texmf/tex/aa-package/bibtex/aa}
\bibliography{/home/tho/texmf/tex/aa-package/bibtex/thbib}

\begin{thebibliography}{56}
\expandafter\ifx\csname natexlab\endcsname\relax\def\natexlab#1{#1}\fi

\bibitem[{{Agudo} {et~al.}(2007){Agudo}, {Bach}, {Krichbaum}, {Marscher},
  {Gonidakis}, {Diamond}, {Perucho}, {Alef}, {Graham}, {Witzel}, {Zensus},
  {Bremer}, {Acosta-Pulido}, \& {Barrena}}]{agudo07}
{Agudo}, I., {Bach}, U., {Krichbaum}, T.~P., {et~al.} 2007, \aap, 476, L17

\bibitem[{{Alberdi} {et~al.}(2000){Alberdi}, {G{\'o}mez}, {Marcaide},
  {Marscher}, \& {P{\'e}rez-Torres}}]{alberdi00}
{Alberdi}, A., {G{\'o}mez}, J.~L., {Marcaide}, J.~M., {Marscher}, A.~P., \&
  {P{\'e}rez-Torres}, M.~A. 2000, \aap, 361, 529

\bibitem[{{Alberdi} {et~al.}(1993){Alberdi}, {Marcaide}, {Marscher}, {Zhang},
  {Elosegui}, {Gomez}, \& {Shaffer}}]{alberdi93}
{Alberdi}, A., {Marcaide}, J.~M., {Marscher}, A.~P., {et~al.} 1993, \apj, 402,
  160

\bibitem[{{Angel} \& {Stockman}(1980)}]{angel80}
{Angel}, J.~R.~P. \& {Stockman}, H.~S. 1980, \araa, 18, 321

\bibitem[{{Barthel}(1989)}]{barthel89}
{Barthel}, P.~D. 1989, \apj, 336, 606

\bibitem[{{Britzen} {et~al.}(2007){Britzen}, {Brinkmann}, {Campbell},
  {Gliozzi}, {Readhead}, {Browne}, \& {Wilkinson}}]{britzen07}
{Britzen}, S., {Brinkmann}, W., {Campbell}, R.~M., {et~al.} 2007, \aap, 476,
  759

\bibitem[{{Britzen} {et~al.}(2008){Britzen}, {Vermeulen}, {Campbell}, {Taylor},
  {Pearson}, {Readhead}, {Xu}, {Browne}, {Henstock}, \&
  {Wilkinson}}]{britzen08}
{Britzen}, S., {Vermeulen}, R.~C., {Campbell}, R.~M., {et~al.} 2008, \aap, 484,
  119

\bibitem[{{Cara} \& {Lister}(2008)}]{cara08}
{Cara}, M. \& {Lister}, M.~L. 2008, \apj, 674, 111

\bibitem[{{Cohen} {et~al.}(2007){Cohen}, {Lister}, {Homan}, {Kadler},
  {Kellermann}, {Kovalev}, \& {Vermeulen}}]{cohen07}
{Cohen}, M.~H., {Lister}, M.~L., {Homan}, D.~C., {et~al.} 2007, \apj, 658, 232

\bibitem[{{Fey} {et~al.}(1997){Fey}, {Eubanks}, \& {Kingham}}]{fey97}
{Fey}, A.~L., {Eubanks}, M., \& {Kingham}, K.~A. 1997, \aj, 114, 2284

\bibitem[{{Ghisellini} {et~al.}(1993){Ghisellini}, {Padovani}, {Celotti}, \&
  {Maraschi}}]{ghisellini93}
{Ghisellini}, G., {Padovani}, P., {Celotti}, A., \& {Maraschi}, L. 1993, \apj,
  407, 65

\bibitem[{{Guerra} \& {Daly}(1997)}]{guerra97}
{Guerra}, E.~J. \& {Daly}, R.~A. 1997, \apj, 491, 483

\bibitem[{{Guijosa} \& {Daly}(1996)}]{guijosa96}
{Guijosa}, A. \& {Daly}, R.~A. 1996, \apj, 461, 600

\bibitem[{{Homan} {et~al.}(2006){Homan}, {Kovalev}, {Lister}, {Ros},
  {Kellermann}, {Cohen}, {Vermeulen}, {Zensus}, \& {Kadler}}]{homan06}
{Homan}, D.~C., {Kovalev}, Y.~Y., {Lister}, M.~L., {et~al.} 2006, \apjl, 642,
  L115

\bibitem[{{Homan} {et~al.}(2001){Homan}, {Ojha}, {Wardle}, {Roberts}, {Aller},
  {Aller}, \& {Hughes}}]{homan01}
{Homan}, D.~C., {Ojha}, R., {Wardle}, J.~F.~C., {et~al.} 2001, \apj, 549, 840

\bibitem[{{Hovatta} {et~al.}(2008{\natexlab{a}}){Hovatta}, {Lehto}, \&
  {Tornikoski}}]{hovatta08b}
{Hovatta}, T., {Lehto}, H.~J., \& {Tornikoski}, M. 2008{\natexlab{a}}, \aap,
  488, 897

\bibitem[{{Hovatta} {et~al.}(2008{\natexlab{b}}){Hovatta}, {Nieppola},
  {Tornikoski}, {Valtaoja}, {Aller}, \& {Aller}}]{hovatta08}
{Hovatta}, T., {Nieppola}, E., {Tornikoski}, M., {et~al.} 2008{\natexlab{b}},
  \aap, 485, 51

\bibitem[{Hovatta {et~al.}(2007)Hovatta, Tornikoski, Lainela, Lehto, Valtaoja,
  Torniainen, Aller, \& Aller}]{hovatta07}
Hovatta, T., Tornikoski, M., Lainela, M., {et~al.} 2007, \aap, 469, 899

\bibitem[{{Hughes} {et~al.}(2002){Hughes}, {Miller}, \& {Duncan}}]{hughes02}
{Hughes}, P.~A., {Miller}, M.~A., \& {Duncan}, G.~C. 2002, \apj, 572, 713

\bibitem[{{Impey} {et~al.}(2000){Impey}, {Bychkov}, {Tapia}, {Gnedin}, \&
  {Pustilnik}}]{impey00}
{Impey}, C.~D., {Bychkov}, V., {Tapia}, S., {Gnedin}, Y., \& {Pustilnik}, S.
  2000, \aj, 119, 1542

\bibitem[{{Impey} {et~al.}(1991){Impey}, {Lawrence}, \& {Tapia}}]{impey91}
{Impey}, C.~D., {Lawrence}, C.~R., \& {Tapia}, S. 1991, \apj, 375, 46

\bibitem[{{Impey} \& {Tapia}(1990)}]{impey90}
{Impey}, C.~D. \& {Tapia}, S. 1990, \apj, 354, 124

\bibitem[{{Jorstad} {et~al.}(2005){Jorstad}, {Marscher}, {Lister}, {Stirling},
  {Cawthorne}, {Gear}, {G{\'o}mez}, {Stevens}, {Smith}, {Forster}, \&
  {Robson}}]{jorstad05}
{Jorstad}, S.~G., {Marscher}, A.~P., {Lister}, M.~L., {et~al.} 2005, \aj, 130,
  1418

\bibitem[{{Jorstad} {et~al.}(2001){Jorstad}, {Marscher}, {Mattox}, {Wehrle},
  {Bloom}, \& {Yurchenko}}]{jorstad01}
{Jorstad}, S.~G., {Marscher}, A.~P., {Mattox}, J.~R., {et~al.} 2001, \apjs,
  134, 181

\bibitem[{{Jorstad} {et~al.}(2007){Jorstad}, {Marscher}, {Stevens}, {Smith},
  {Forster}, {Gear}, {Cawthorne}, {Lister}, {Stirling}, {G{\'o}mez}, {Greaves},
  \& {Robson}}]{jorstad07}
{Jorstad}, S.~G., {Marscher}, A.~P., {Stevens}, J.~A., {et~al.} 2007, \aj, 134,
  799

\bibitem[{{Kellermann} {et~al.}(2004){Kellermann}, {Lister}, {Homan},
  {Vermeulen}, {Cohen}, {Ros}, {Kadler}, {Zensus}, \& {Kovalev}}]{kellermann04}
{Kellermann}, K.~I., {Lister}, M.~L., {Homan}, D.~C., {et~al.} 2004, \apj, 609,
  539

\bibitem[{{Kovalev} {et~al.}(2005){Kovalev}, {Kellermann}, {Lister}, {Homan},
  {Vermeulen}, {Cohen}, {Ros}, {Kadler}, {Lobanov}, {Zensus}, {Kardashev},
  {Gurvits}, {Aller}, \& {Aller}}]{kovalev05}
{Kovalev}, Y.~Y., {Kellermann}, K.~I., {Lister}, M.~L., {et~al.} 2005, \aj,
  130, 2473

\bibitem[{L\"ahteenm\"aki \& Valtaoja(1999)}]{lahteenmaki99b}
L\"ahteenm\"aki, A. \& Valtaoja, E. 1999, \apj, 521, 493, LV99

\bibitem[{L\"ahteenm\"aki {et~al.}(1999)L\"ahteenm\"aki, Valtaoja, \&
  Wiik}]{lahteenmaki99}
L\"ahteenm\"aki, A., Valtaoja, E., \& Wiik, K. 1999, \apj, 511, 112

\bibitem[{{Lister} \& {Smith}(2000)}]{lister00}
{Lister}, M.~L. \& {Smith}, P.~S. 2000, \apj, 541, 66

\bibitem[{{Liu} \& {Zhang}(2007)}]{liu07}
{Liu}, Y. \& {Zhang}, S.~N. 2007, \apj, 667, 724

\bibitem[{{Moore} \& {Stockman}(1984)}]{moore84}
{Moore}, R.~L. \& {Stockman}, H.~S. 1984, \apj, 279, 465

\bibitem[{Nieppola {et~al.}(2007)Nieppola, Tornikoski, L\"ahteenm\"aki,
  Valtaoja, Hakala, Hovatta, Kotiranta, Nummila, Ojala, Parviainen, Ranta,
  Saloranta, Torniainen, \& Tr\"oller}]{nieppola07}
Nieppola, E., Tornikoski, M., L\"ahteenm\"aki, A., {et~al.} 2007, \aj, 133,
  1947

\bibitem[{{Padovani} \& {Urry}(1992)}]{padovani92}
{Padovani}, P. \& {Urry}, C.~M. 1992, \apj, 387, 449

\bibitem[{{Piner} {et~al.}(2007){Piner}, {Mahmud}, {Fey}, \&
  {Gospodinova}}]{piner07}
{Piner}, B.~G., {Mahmud}, M., {Fey}, A.~L., \& {Gospodinova}, K. 2007, \aj,
  133, 2357

\bibitem[{{Punsly}(1996)}]{punsly96}
{Punsly}, B. 1996, \apj, 473, 152

\bibitem[{{Readhead}(1994)}]{readhead94}
{Readhead}, A.~C.~S. 1994, \apj, 426, 51

\bibitem[{Salonen {et~al.}(1987)Salonen, Ter\"asranta, Urpo, Tiuri, Moiseev,
  Nesterov, Valtaoja, Haarala, Lehto, Valtaoja, Teerikorpi, \&
  Valtonen}]{salonen87}
Salonen, E., Ter\"asranta, H., Urpo, S., {et~al.} 1987, \aaps, 70, 409

\bibitem[{{Savolainen} {et~al.}(2002){Savolainen}, {Wiik}, {Valtaoja},
  {Jorstad}, \& {Marscher}}]{savolainen02}
{Savolainen}, T., {Wiik}, K., {Valtaoja}, E., {Jorstad}, S.~G., \& {Marscher},
  A.~P. 2002, \aap, 394, 851

\bibitem[{{Savolainen} {et~al.}(2006){Savolainen}, {Wiik}, {Valtaoja}, \&
  {Tornikoski}}]{savolainen06b}
{Savolainen}, T., {Wiik}, K., {Valtaoja}, E., \& {Tornikoski}, M. 2006, \aap,
  446, 71

\bibitem[{{Schmitt} {et~al.}(2001){Schmitt}, {Antonucci}, {Ulvestad}, {Kinney},
  {Clarke}, \& {Pringle}}]{schmitt01}
{Schmitt}, H.~R., {Antonucci}, R.~R.~J., {Ulvestad}, J.~S., {et~al.} 2001,
  \apj, 555, 663

\bibitem[{{Sluse} {et~al.}(2005){Sluse}, {Hutsem{\'e}kers}, {Lamy}, {Cabanac},
  \& {Quintana}}]{sluse05}
{Sluse}, D., {Hutsem{\'e}kers}, D., {Lamy}, H., {Cabanac}, R., \& {Quintana},
  H. 2005, \aap, 433, 757

\bibitem[{{Smith} {et~al.}(1985){Smith}, {Balonek}, {Heckert}, {Elston}, \&
  {Schmidt}}]{smith85}
{Smith}, P.~S., {Balonek}, T.~J., {Heckert}, P.~A., {Elston}, R., \& {Schmidt},
  G.~D. 1985, \aj, 90, 1184

\bibitem[{{Stickel} \& {Kuehr}(1994)}]{stickel94}
{Stickel}, M. \& {Kuehr}, H. 1994, \aaps, 105, 67

\bibitem[{{Taylor} {et~al.}(1996){Taylor}, {Vermeulen}, {Readhead}, {Pearson},
  {Henstock}, \& {Wilkinson}}]{taylor96}
{Taylor}, G.~B., {Vermeulen}, R.~C., {Readhead}, A.~C.~S., {et~al.} 1996,
  \apjs, 107, 37

\bibitem[{Ter\"asranta {et~al.}(2004)Ter\"asranta, Achren, Hanski, Heikkil\"a,
  Holopainen, Joutsamo, Juhola, Karlamaa, Katajainen, Kein\"anen, Koivisto,
  Koskimies, K\"on\"onen, Lainela, L\"ahteenm\"aki, M\"akinen, Niemel\"a,
  Nurmi, Pursimo, Rekola, Savolainen, Tornikoski, Torppa, Valtonen, Varjonen,
  Vilenius, Virtanen, \& Wiren}]{terasranta04}
Ter\"asranta, H., Achren, J., Hanski, M., {et~al.} 2004, \aap, 427, 769

\bibitem[{Ter\"asranta {et~al.}(1998)Ter\"asranta, Tornikoski, Mujunen,
  Karlamaa, Valtonen, Henelius, Urpo, Lainela, Pursimo, Nilsson, Wiren,
  L\"ahteenm\"aki, Korpi, Rekola, Hein\"am\"aki, Hanski, Nurmi, Kokkonen,
  Kein\"anen, Joutsamo, Oksanen, Pietil\"a, Valtaoja, Valtonen, \&
  K\"on\"onen}]{terasranta98}
Ter\"asranta, H., Tornikoski, M., Mujunen, A., {et~al.} 1998, \aaps, 132, 305

\bibitem[{Ter\"asranta {et~al.}(1992)Ter\"asranta, Tornikoski, Valtaoja, Urpo,
  Nesterov, Lainela, Kotilainen, Wiren, Laine, Nilsson, \&
  Valtonen}]{terasranta92}
Ter\"asranta, H., Tornikoski, M., Valtaoja, E., {et~al.} 1992, \aaps, 94, 121

\bibitem[{Ter\"asranta {et~al.}(2005)Ter\"asranta, Wiren, Koivisto, Saarinen,
  \& Hovatta}]{terasranta05}
Ter\"asranta, H., Wiren, S., Koivisto, P., Saarinen, V., \& Hovatta, T. 2005,
  \aap, 440, 409

\bibitem[{{Urry} \& {Padovani}(1995)}]{urry95}
{Urry}, C.~M. \& {Padovani}, P. 1995, \pasp, 107, 803

\bibitem[{{Urry} {et~al.}(1991){Urry}, {Padovani}, \& {Stickel}}]{urry91}
{Urry}, C.~M., {Padovani}, P., \& {Stickel}, M. 1991, \apj, 382, 501

\bibitem[{Valtaoja {et~al.}(1999)Valtaoja, L\"ahteenm\"aki, Ter\"asranta, \&
  Lainela}]{valtaoja99}
Valtaoja, E., L\"ahteenm\"aki, A., Ter\"asranta, H., \& Lainela, M. 1999,
  \apjs, 120, 95

\bibitem[{{Vermeulen} \& {Cohen}(1994)}]{vermeulen94}
{Vermeulen}, R.~C. \& {Cohen}, M.~H. 1994, \apj, 430, 467

\bibitem[{{V{\'e}ron-Cetty} \& {V{\'e}ron}(2006)}]{veron06}
{V{\'e}ron-Cetty}, M.-P. \& {V{\'e}ron}, P. 2006, \aap, 455, 773

\bibitem[{{Wills} {et~al.}(1992){Wills}, {Wills}, {Breger}, {Antonucci}, \&
  {Barvainis}}]{wills92}
{Wills}, B.~J., {Wills}, D., {Breger}, M., {Antonucci}, R.~R.~J., \&
  {Barvainis}, R. 1992, \apj, 398, 454

\bibitem[{{Wright}(2006)}]{wright06}
{Wright}, E.~L. 2006, \pasp, 118, 1711

\end{thebibliography}

\longtab{1}{
\small{
\begin{longtable}{lllcccrrrrrrrc}
\caption[]{\label{table:sourcelist} Doppler boosting factors, Lorentz factor and viewing angles for all sources.}\\
\hline
\hline
B1950-name 	  & 	 Other name 	  & 	 Type 	  & 	 $\nu$	  & 	 $z$ 	  & 	 Quality 	  & 	 $\log(T_b)$ 	  & 	 $D_\mathrm{var}$ 	  & 	 $\beta_\mathrm{app}$ 	  & 	 $\Gamma_\mathrm{var}$ 	  & 	 $\theta_\mathrm{var}$ 	  & 	 $\log{R}$ 	  & 	 $P_\mathrm{max}$ 	  & 	 ref. 	  \\
\hline 
\endfirsthead 
\caption{continued.}\\
B1950-name 	  & 	 Other name 	  & 	 Type 	  & 	 $\nu$ 	  & 	 $z$ 	  & 	 Quality 	  & 	 $\log(T_b)$ 	  & 	 $D_\mathrm{var}$ 	  & 	 $\beta_\mathrm{app}$ 	  & 	 $\Gamma_\mathrm{var}$ 	  & 	 $\theta_\mathrm{var}$ 	  & 	 $\log{R}$ 	  & 	 $P_\mathrm{max}$ 	  & 	 ref. 	  \\
\hline
\endhead
\hline
\endfoot
\object{0003$-$066} 	  & 	 NRAO 5 	  & 	 BLO 	  & 	 22 	  & 	 0.347 	  & 	 A 	  & 	 12.82 	  & 	 5.1 	  & 	 2.660 	  & 	 3.3 	  & 	 9.5 	  & 	 -0.133 	  & 	 3.4 	  & 	 11 	  \\
 \object{0007+106} 	  & 	 III ZW 2 	  & 	 GAL 	  & 	 37 	  & 	 0.089 	  & 	 E 	  & 	 11.37 	  & 	 1.7 	  & 	 0.98 	  & 	 1.4 	  & 	 35.4 	  & 	 0.004 	  & 	 0.2 	  & 	 12 	  \\
 \object{0016+731} 	  & 	  	  & 	 LPQ 	  & 	 37 	  & 	 1.781 	  & 	 A 	  & 	 13.40 	  & 	 7.9 	  & 	 6.67 	  & 	 6.8 	  & 	 7.1 	  & 	 -0.104 	  & 	 0.6 	  & 	 11 	  \\
 \object{0048$-$097} 	  & 	 PKS 0048-097 	  & 	 BLO 	  & 	 22 	  & 	 0.300 	  & 	 A 	  & 	 13.65 	  & 	 9.6 	  & 	 ... 	  & 	 ... 	  & 	 ... 	  & 	 -0.082 	  & 	 27.1 	  & 	 2 	  \\
 \object{0059+581} 	  & 	  	  & 	 QSO 	  & 	 22 	  & 	 0.644 	  & 	 G 	  & 	 14.15 	  & 	 14.1 	  & 	 11.1 	  & 	 11.5 	  & 	 4.0 	  & 	 0.018 	  & 	 ... 	  & 	 ... 	  \\
 \object{0106+013} 	  & 	 OC 012 	  & 	 HPQ 	  & 	 22 	  & 	 2.099 	  & 	 E 	  & 	 14.50 	  & 	 18.4 	  & 	 26.16 	  & 	 27.8 	  & 	 2.9 	  & 	 -0.172 	  & 	 7.1 	  & 	 11 	  \\
 \object{0133+476} 	  & 	 DA 55 	  & 	 HPQ 	  & 	 22 	  & 	 0.859 	  & 	 G 	  & 	 14.65 	  & 	 20.7 	  & 	 12.95 	  & 	 14.4 	  & 	 2.5 	  & 	 -0.068 	  & 	 20.8 	  & 	 3 	  \\
 \object{0149+218} 	  & 	  	  & 	 LPQ 	  & 	 37 	  & 	 1.320 	  & 	 A 	  & 	 12.83 	  & 	 5.1 	  & 	 ... 	  & 	 ... 	  & 	 ... 	  & 	 -0.024 	  & 	 ... 	  & 	 ... 	  \\
 \object{0202+149} 	  & 	 4C 15.05 	  & 	 HPQ 	  & 	 22 	  & 	 0.405 	  & 	 G 	  & 	 14.24 	  & 	 15.1 	  & 	 8.31 	  & 	 9.9 	  & 	 3.2 	  & 	 -0.216 	  & 	 4.0 	  & 	 2 	  \\
 \object{0212+735} 	  & 	  	  & 	 HPQ 	  & 	 22 	  & 	 2.367 	  & 	 A 	  & 	 13.48 	  & 	 8.5 	  & 	 7.35 	  & 	 7.5 	  & 	 6.7 	  & 	 -0.268 	  & 	 7.8 	  & 	 2 	  \\
 \object{0219+428} 	  & 	 3C 66A 	  & 	 BLO 	  & 	 37 	  & 	 0.444 	  & 	 A 	  & 	 11.92 	  & 	 2.6 	  & 	 ... 	  & 	 ... 	  & 	 ... 	  & 	 ... 	  & 	 29.7 	  & 	 5 	  \\
 \object{0224+671} 	  & 	  	  & 	 QSO 	  & 	 22 	  & 	 0.523 	  & 	 A 	  & 	 13.43 	  & 	 8.2 	  & 	 11.67 	  & 	 12.5 	  & 	 6.6 	  & 	 -0.308 	  & 	 ... 	  & 	 ... 	  \\
 \object{0234+285} 	  & 	 4C 28.07 	  & 	 HPQ 	  & 	 37 	  & 	 1.213 	  & 	 G 	  & 	 14.32 	  & 	 16.1 	  & 	 12.23 	  & 	 12.7 	  & 	 3.4 	  & 	 -0.090 	  & 	 11.3 	  & 	 2 	  \\
 \object{0235+164} 	  & 	  	  & 	 BLO 	  & 	 22 	  & 	 0.940 	  & 	 G 	  & 	 14.84 	  & 	 24.0 	  & 	 2.000 	  & 	 12.1 	  & 	 0.4 	  & 	 -0.092 	  & 	 44.7 	  & 	 2 	  \\
 \object{0306+102} 	  & 	 PKS 0306+102 	  & 	 BLO 	  & 	 22 	  & 	 0.863 	  & 	 A 	  & 	 13.97 	  & 	 12.3 	  & 	 ... 	  & 	 ... 	  & 	 ... 	  & 	 ... 	  & 	 ... 	  & 	 ... 	  \\
 \object{0316+413} 	  & 	 3C 84 	  & 	 GAL 	  & 	 37 	  & 	 0.018 	  & 	 G 	  & 	 9.26 	  & 	 0.3 	  & 	 0.32 	  & 	 1.8 	  & 	 39.1 	  & 	 -0.654 	  & 	 6.0 	  & 	 11 	  \\
 \object{0333+321} 	  & 	 NRAO 140 	  & 	 LPQ 	  & 	 22 	  & 	 1.263 	  & 	 G 	  & 	 14.74 	  & 	 22.2 	  & 	 12.7 	  & 	 14.7 	  & 	 2.2 	  & 	 -0.140 	  & 	 0.7 	  & 	 12 	  \\
 \object{0336$-$019} 	  & 	 CTA 026 	  & 	 HPQ 	  & 	 22 	  & 	 0.852 	  & 	 A 	  & 	 14.42 	  & 	 17.4 	  & 	 22.32 	  & 	 23.0 	  & 	 3.2 	  & 	 -0.133 	  & 	 19.4 	  & 	 2 	  \\
 \object{0355+508} 	  & 	 NRAO 150 	  & 	 QSO 	  & 	 37 	  & 	 1.510 	  & 	 G 	  & 	 14.05 	  & 	 13.1 	  & 	 ... 	  & 	 ... 	  & 	 ... 	  & 	 -0.015 	  & 	 ... 	  & 	 ... 	  \\
 \object{0415+379} 	  & 	 3C 111 	  & 	 GAL 	  & 	 37 	  & 	 0.049 	  & 	 G 	  & 	 12.09 	  & 	 2.9 	  & 	 5.94 	  & 	 7.7 	  & 	 15.5 	  & 	 -0.432 	  & 	 3.1 	  & 	 5 	  \\
 \object{0420$-$014} 	  & 	 OA 129 	  & 	 HPQ 	  & 	 37 	  & 	 0.915 	  & 	 E 	  & 	 14.59 	  & 	 19.9 	  & 	 7.52 	  & 	 11.4 	  & 	 1.9 	  & 	 -0.097 	  & 	 26.2 	  & 	 11 	  \\
 \object{0430+052} 	  & 	 3C 120 	  & 	 GAL 	  & 	 22 	  & 	 0.033 	  & 	 G 	  & 	 13.01 	  & 	 5.9 	  & 	 5.42 	  & 	 5.5 	  & 	 9.7 	  & 	 -0.488 	  & 	 0.5 	  & 	 5 	  \\
 \object{0440$-$003} 	  & 	 NRAO 190 	  & 	 HPQ 	  & 	 22 	  & 	 0.844 	  & 	 A 	  & 	 14.03 	  & 	 12.9 	  & 	 ... 	  & 	 ... 	  & 	 ... 	  & 	 -0.291 	  & 	 12.5 	  & 	 11 	  \\
 \object{0458$-$020} 	  & 	 PKS 0458-020 	  & 	 HPQ 	  & 	 37 	  & 	 2.286 	  & 	 G 	  & 	 14.29 	  & 	 15.8 	  & 	 16.2 	  & 	 16.2 	  & 	 3.6 	  & 	 -0.134 	  & 	 17.3 	  & 	 2 	  \\
 \object{0528+134} 	  & 	 PKS 0528+134 	  & 	 HPQ 	  & 	 22 	  & 	 2.070 	  & 	 E 	  & 	 15.18 	  & 	 31.2 	  & 	 18.73 	  & 	 21.2 	  & 	 1.6 	  & 	 -0.059 	  & 	 4.1 	  & 	 5 	  \\
 \object{0552+398} 	  & 	 DA 193 	  & 	 LPQ 	  & 	 37 	  & 	 2.363 	  & 	 A 	  & 	 14.91 	  & 	 25.2 	  & 	 0.45 	  & 	 12.6 	  & 	 0.1 	  & 	 -0.157 	  & 	 1.9 	  & 	 12 	  \\
 \object{0605$-$085} 	  & 	 PKS 0605-085 	  & 	 HPQ 	  & 	 37 	  & 	 0.872 	  & 	 A 	  & 	 13.34 	  & 	 7.6 	  & 	 19.98 	  & 	 30.2 	  & 	 5.0 	  & 	 -0.191 	  & 	 9.8 	  & 	 11 	  \\
 \object{0642+449} 	  & 	 OH 471 	  & 	 LPQ 	  & 	 22 	  & 	 3.396 	  & 	 G 	  & 	 13.78 	  & 	 10.7 	  & 	 0.53 	  & 	 5.4 	  & 	 0.5 	  & 	 -0.192 	  & 	 1.7 	  & 	 7 	  \\
 \object{0716+714} 	  & 	  	  & 	 BLO 	  & 	 22 	  & 	 0.310 	  & 	 E 	  & 	 13.81 	  & 	 10.9 	  & 	 10.22 	  & 	 10.3 	  & 	 5.2 	  & 	 -0.219 	  & 	 17.3 	  & 	 4 	  \\
 \object{0735+178} 	  & 	 PKS 0735+17 	  & 	 BLO 	  & 	 37 	  & 	 0.424 	  & 	 G 	  & 	 12.43 	  & 	 3.8 	  & 	 ... 	  & 	 ... 	  & 	 ... 	  & 	 -0.378 	  & 	 36.0 	  & 	 2 	  \\
 \object{0736+017} 	  & 	  	  & 	 HPQ 	  & 	 22 	  & 	 0.191 	  & 	 E 	  & 	 13.50 	  & 	 8.6 	  & 	 14.73 	  & 	 17.0 	  & 	 5.8 	  & 	 -0.157 	  & 	 8.4 	  & 	 11 	  \\
 \object{0754+100} 	  & 	 OI 090.4 	  & 	 BLO 	  & 	 37 	  & 	 0.266 	  & 	 A 	  & 	 12.95 	  & 	 5.6 	  & 	 14.530 	  & 	 21.7 	  & 	 6.9 	  & 	 -0.168 	  & 	 26.0 	  & 	 1 	  \\
 \object{0804+499} 	  & 	  	  & 	 HPQ 	  & 	 22 	  & 	 1.436 	  & 	 E 	  & 	 15.35 	  & 	 35.5 	  & 	 1.81 	  & 	 17.8 	  & 	 0.2 	  & 	 -0.096 	  & 	 8.6 	  & 	 2 	  \\
 \object{0814+425} 	  & 	  	  & 	 BLO 	  & 	 22 	  & 	 0.245 	  & 	 G 	  & 	 12.69 	  & 	 4.6 	  & 	 1.720 	  & 	 2.7 	  & 	 8.4 	  & 	 -0.139 	  & 	 8.7 	  & 	 2 	  \\
 \object{0827+243} 	  & 	 OJ 248 	  & 	 LPQ 	  & 	 37 	  & 	 0.941 	  & 	 A 	  & 	 14.05 	  & 	 13.1 	  & 	 21.3 	  & 	 23.9 	  & 	 3.9 	  & 	 -0.037 	  & 	 1.5 	  & 	 7 	  \\
 \object{0836+710} 	  & 	 4C 71.07 	  & 	 LPQ 	  & 	 37 	  & 	 2.218 	  & 	 A 	  & 	 14.33 	  & 	 16.3 	  & 	 25.39 	  & 	 28.0 	  & 	 3.2 	  & 	 -0.119 	  & 	 1.1 	  & 	 11 	  \\
 \object{0847$-$120} 	  & 	 J 0850-1213 	  & 	 QSO 	  & 	 37 	  & 	 0.566 	  & 	 G 	  & 	 14.35 	  & 	 16.5 	  & 	 ... 	  & 	 ... 	  & 	 ... 	  & 	 ... 	  & 	 ... 	  & 	 ... 	  \\
 \object{0851+202} 	  & 	 OJ 287 	  & 	 BLO 	  & 	 22 	  & 	 0.306 	  & 	 E 	  & 	 14.39 	  & 	 17.0 	  & 	 15.31 	  & 	 15.4 	  & 	 3.3 	  & 	 -0.181 	  & 	 15.6 	  & 	 5 	  \\
 \object{0923+392} 	  & 	 4C 39.25 	  & 	 LPQ 	  & 	 37 	  & 	 0.695 	  & 	 G 	  & 	 12.60 	  & 	 4.3 	  & 	 42.94 	  & 	 216.1 	  & 	 2.6 	  & 	 -1.757 	  & 	 0.5 	  & 	 2 	  \\
 \object{0945+408} 	  & 	 4C 40.24 	  & 	 HPQ 	  & 	 37 	  & 	 1.249 	  & 	 A 	  & 	 13.13 	  & 	 6.4 	  & 	 18.47 	  & 	 29.8 	  & 	 5.5 	  & 	 -0.274 	  & 	 6.5 	  & 	 3 	  \\
 \object{0953+254} 	  & 	  	  & 	 LPQ 	  & 	 37 	  & 	 0.712 	  & 	 A 	  & 	 12.61 	  & 	 4.3 	  & 	 ... 	  & 	 ... 	  & 	 ... 	  & 	 -0.268 	  & 	 2.2 	  & 	 2 	  \\
 \object{0954+658} 	  & 	 S4 0954+65 	  & 	 BLO 	  & 	 37 	  & 	 0.367 	  & 	 A 	  & 	 13.07 	  & 	 6.2 	  & 	 ... 	  & 	 ... 	  & 	 ... 	  & 	 -0.054 	  & 	 33.7 	  & 	 11 	  \\
 \object{1055+018} 	  & 	 OL 093 	  & 	 HPQ 	  & 	 37 	  & 	 0.888 	  & 	 G 	  & 	 13.96 	  & 	 12.2 	  & 	 10.96 	  & 	 11.1 	  & 	 4.7 	  & 	 -0.159 	  & 	 5.0 	  & 	 2 	  \\
 \object{1156+295} 	  & 	 4C 29.45 	  & 	 HPQ 	  & 	 22 	  & 	 0.729 	  & 	 E 	  & 	 15.06 	  & 	 28.5 	  & 	 24.85 	  & 	 25.1 	  & 	 2.0 	  & 	 -0.080 	  & 	 9.2 	  & 	 12 	  \\
 \object{1219+285} 	  & 	 ON 231 	  & 	 BLO 	  & 	 22 	  & 	 0.102 	  & 	 A 	  & 	 10.96 	  & 	 1.2 	  & 	 ... 	  & 	 ... 	  & 	 ... 	  & 	 -0.363 	  & 	 10.0 	  & 	 1 	  \\
 \object{1222+216} 	  & 	 PKS 1222+216 	  & 	 LPQ 	  & 	 37 	  & 	 0.432 	  & 	 A 	  & 	 12.85 	  & 	 5.2 	  & 	 21.12 	  & 	 45.5 	  & 	 5.1 	  & 	 -0.210 	  & 	 1.5 	  & 	 9 	  \\
 \object{1226+023} 	  & 	 3C 273 	  & 	 LPQ 	  & 	 22 	  & 	 0.158 	  & 	 E 	  & 	 14.39 	  & 	 17.0 	  & 	 13.59 	  & 	 14.0 	  & 	 3.3 	  & 	 -0.809 	  & 	 0.5 	  & 	 2 	  \\
 \object{1253$-$055} 	  & 	 3C 279 	  & 	 HPQ 	  & 	 37 	  & 	 0.536 	  & 	 E 	  & 	 14.84 	  & 	 24.0 	  & 	 20.64 	  & 	 20.9 	  & 	 2.4 	  & 	 -0.242 	  & 	 39.2 	  & 	 5 	  \\
 \object{1308+326} 	  & 	 AU CV n 	  & 	 HPQ 	  & 	 22 	  & 	 0.997 	  & 	 G 	  & 	 14.26 	  & 	 15.4 	  & 	 27.500 	  & 	 32.2 	  & 	 3.2 	  & 	 -0.096 	  & 	 25.0 	  & 	 2 	  \\
 \object{1324+224} 	  & 	  	  & 	 QSO 	  & 	 22 	  & 	 1.400 	  & 	 A 	  & 	 14.68 	  & 	 21.2 	  & 	 3.04 	  & 	 10.9 	  & 	 0.8 	  & 	 -0.075 	  & 	 ... 	  & 	 ... 	  \\
 \object{1413+135} 	  & 	  	  & 	 BLO 	  & 	 22 	  & 	 0.247 	  & 	 G 	  & 	 13.96 	  & 	 12.2 	  & 	 1.860 	  & 	 6.3 	  & 	 1.4 	  & 	 -0.112 	  & 	 13.0 	  & 	 8 	  \\
 \object{1418+546} 	  & 	 OQ 530 	  & 	 BLO 	  & 	 22 	  & 	 0.151 	  & 	 A 	  & 	 12.83 	  & 	 5.1 	  & 	 ... 	  & 	 ... 	  & 	 ... 	  & 	 -0.175 	  & 	 19.0 	  & 	 1 	  \\
 \object{1502+106} 	  & 	 OR 103 	  & 	 HPQ 	  & 	 37 	  & 	 1.839 	  & 	 E 	  & 	 13.94 	  & 	 12.0 	  & 	 14.6 	  & 	 14.9 	  & 	 4.7 	  & 	 -0.202 	  & 	 3.0 	  & 	 2 	  \\
 \object{1510$-$089} 	  & 	 PKS 1510-089 	  & 	 HPQ 	  & 	 37 	  & 	 0.360 	  & 	 E 	  & 	 14.37 	  & 	 16.7 	  & 	 20.29 	  & 	 20.7 	  & 	 3.4 	  & 	 -0.167 	  & 	 7.8 	  & 	 11 	  \\
 \object{1538+149} 	  & 	 4C 14.60 	  & 	 BLO 	  & 	 37 	  & 	 0.605 	  & 	 A 	  & 	 12.60 	  & 	 4.3 	  & 	 8.750 	  & 	 11.2 	  & 	 10.5 	  & 	 -0.296 	  & 	 17.4 	  & 	 2 	  \\
 \object{1606+106} 	  & 	 4C 10.45 	  & 	 LPQ 	  & 	 22 	  & 	 1.226 	  & 	 G 	  & 	 14.89 	  & 	 25.0 	  & 	 18.78 	  & 	 19.6 	  & 	 2.2 	  & 	 -0.081 	  & 	 1.9 	  & 	 11 	  \\
 \object{1611+343} 	  & 	 DA 406 	  & 	 LPQ 	  & 	 37 	  & 	 1.401 	  & 	 A 	  & 	 14.11 	  & 	 13.7 	  & 	 14.17 	  & 	 14.2 	  & 	 4.2 	  & 	 -0.179 	  & 	 2.3 	  & 	 6 	  \\
 \object{1633+382} 	  & 	 4C 38.41 	  & 	 HPQ 	  & 	 22 	  & 	 1.814 	  & 	 E 	  & 	 14.70 	  & 	 21.5 	  & 	 29.13 	  & 	 30.5 	  & 	 2.5 	  & 	 -0.256 	  & 	 7.0 	  & 	 6 	  \\
 \object{1637+574} 	  & 	 OS 562 	  & 	 LPQ 	  & 	 22 	  & 	 0.751 	  & 	 A 	  & 	 14.13 	  & 	 14.0 	  & 	 10.61 	  & 	 11.0 	  & 	 4.0 	  & 	 -0.093 	  & 	 1.5 	  & 	 12 	  \\
 \object{1641+399} 	  & 	 3C 345 	  & 	 HPQ 	  & 	 37 	  & 	 0.593 	  & 	 E 	  & 	 13.38 	  & 	 7.8 	  & 	 19.27 	  & 	 27.7 	  & 	 5.1 	  & 	 -0.412 	  & 	 38.3 	  & 	 5 	  \\
 \object{1725+044} 	  & 	 PKS 1725+044 	  & 	 QSO 	  & 	 22 	  & 	 0.293 	  & 	 A 	  & 	 12.44 	  & 	 3.8 	  & 	 ... 	  & 	 ... 	  & 	 ... 	  & 	 ... 	  & 	 ... 	  & 	 ... 	  \\
 \object{1730$-$130} 	  & 	 NRAO 530 	  & 	 QSO 	  & 	 37 	  & 	 0.902 	  & 	 G 	  & 	 13.79 	  & 	 10.7 	  & 	 35.6 	  & 	 64.6 	  & 	 3.0 	  & 	 -0.291 	  & 	 ... 	  & 	 ... 	  \\
 \object{1739+522} 	  & 	 S4 1739+52 	  & 	 HPQ 	  & 	 22 	  & 	 1.375 	  & 	 E 	  & 	 14.97 	  & 	 26.5 	  & 	 ... 	  & 	 ... 	  & 	 ... 	  & 	 -0.040 	  & 	 3.7 	  & 	 11 	  \\
 \object{1741$-$038} 	  & 	 PKS 1741-038 	  & 	 HPQ 	  & 	 22 	  & 	 1.054 	  & 	 E 	  & 	 14.58 	  & 	 19.7 	  & 	 ... 	  & 	 ... 	  & 	 ... 	  & 	 -0.055 	  & 	 9.2 	  & 	 2 	  \\
 \object{1749+096} 	  & 	 PKS 1749+096 	  & 	 BLO 	  & 	 37 	  & 	 0.322 	  & 	 E 	  & 	 13.93 	  & 	 12.0 	  & 	 5.880 	  & 	 7.5 	  & 	 3.8 	  & 	 -0.030 	  & 	 31.3 	  & 	 2 	  \\
 \object{1803+784} 	  & 	 S5 1803+784 	  & 	 BLO 	  & 	 22 	  & 	 0.684 	  & 	 A 	  & 	 13.96 	  & 	 12.2 	  & 	 8.980 	  & 	 9.5 	  & 	 4.5 	  & 	 -0.250 	  & 	 7.0 	  & 	 3 	  \\
 \object{1807+698} 	  & 	 3C 371.0 	  & 	 BLO 	  & 	 37 	  & 	 0.051 	  & 	 A 	  & 	 10.78 	  & 	 1.1 	  & 	 0.120 	  & 	 1.0 	  & 	 57.3 	  & 	 -0.328 	  & 	 8.0 	  & 	 3 	  \\
 \object{1823+568} 	  & 	 4C 56.27 	  & 	 BLO 	  & 	 22 	  & 	 0.663 	  & 	 A 	  & 	 13.11 	  & 	 6.4 	  & 	 20.950 	  & 	 37.8 	  & 	 5.0 	  & 	 -0.154 	  & 	 29.8 	  & 	 5 	  \\
 \object{1828+487} 	  & 	 3C 380 	  & 	 LPQ 	  & 	 37 	  & 	 0.692 	  & 	 A 	  & 	 12.97 	  & 	 5.7 	  & 	 13.66 	  & 	 19.3 	  & 	 7.1 	  & 	 -0.391 	  & 	 0.4 	  & 	 2 	  \\
 \object{1928+738} 	  & 	 4C 73.18 	  & 	 LPQ 	  & 	 22 	  & 	 0.302 	  & 	 A 	  & 	 11.55 	  & 	 1.9 	  & 	 8.48 	  & 	 19.9 	  & 	 12.8 	  & 	 -0.402 	  & 	 0.8 	  & 	 2 	  \\
 \object{1954+513} 	  & 	  	  & 	 LPQ 	  & 	 22 	  & 	 1.223 	  & 	 A 	  & 	 13.30 	  & 	 7.4 	  & 	 ... 	  & 	 ... 	  & 	 ... 	  & 	 -0.181 	  & 	 1.4 	  & 	 11 	  \\
 \object{2005+403} 	  & 	  	  & 	 QSO 	  & 	 37 	  & 	 1.736 	  & 	 A 	  & 	 14.07 	  & 	 13.3 	  & 	 19.48 	  & 	 21.0 	  & 	 4.0 	  & 	 -0.257 	  & 	 ... 	  & 	 ... 	  \\
 \object{2007+776} 	  & 	 S5 2007+77 	  & 	 BLO 	  & 	 22 	  & 	 0.342 	  & 	 A 	  & 	 13.39 	  & 	 7.9 	  & 	 ... 	  & 	 ... 	  & 	 ... 	  & 	 -0.304 	  & 	 15.1 	  & 	 11 	  \\
 \object{2021+614} 	  & 	 OW 637 	  & 	 GAL 	  & 	 22 	  & 	 0.227 	  & 	 A 	  & 	 11.65 	  & 	 2.1 	  & 	 0.41 	  & 	 1.3 	  & 	 13.3 	  & 	 ... 	  & 	 0.3 	  & 	 3 	  \\
 \object{2022+171} 	  & 	  	  & 	 LPQ 	  & 	 22 	  & 	 1.050 	  & 	 A 	  & 	 12.99 	  & 	 5.8 	  & 	 ... 	  & 	 ... 	  & 	 ... 	  & 	 ... 	  & 	 ... 	  & 	 ... 	  \\
 \object{2121+053} 	  & 	  	  & 	 HPQ 	  & 	 22 	  & 	 1.941 	  & 	 A 	  & 	 14.25 	  & 	 15.3 	  & 	 13.05 	  & 	 13.2 	  & 	 3.7 	  & 	 -0.061 	  & 	 10.7 	  & 	 7 	  \\
 \object{2134+004} 	  & 	 OX 057 	  & 	 LPQ 	  & 	 22 	  & 	 1.932 	  & 	 A 	  & 	 14.32 	  & 	 16.1 	  & 	 5.62 	  & 	 9.0 	  & 	 2.2 	  & 	 -0.405 	  & 	 2.8 	  & 	 12 	  \\
 \object{2136+141} 	  & 	  	  & 	 LPQ 	  & 	 22 	  & 	 2.427 	  & 	 A 	  & 	 13.46 	  & 	 8.3 	  & 	 5.08 	  & 	 5.8 	  & 	 6.2 	  & 	 -0.175 	  & 	 1.6 	  & 	 11 	  \\
 \object{2145+067} 	  & 	  	  & 	 LPQ 	  & 	 37 	  & 	 0.990 	  & 	 G 	  & 	 14.28 	  & 	 15.6 	  & 	 2.49 	  & 	 8.0 	  & 	 1.1 	  & 	 -0.171 	  & 	 0.6 	  & 	 2 	  \\
 \object{2200+420} 	  & 	 BL LAC 	  & 	 BLO 	  & 	 37 	  & 	 0.069 	  & 	 E 	  & 	 13.28 	  & 	 7.3 	  & 	 10.700 	  & 	 11.6 	  & 	 7.3 	  & 	 -0.264 	  & 	 23.0 	  & 	 1 	  \\
 \object{2201+315} 	  & 	 4C 31.63 	  & 	 LPQ 	  & 	 22 	  & 	 0.295 	  & 	 G 	  & 	 13.17 	  & 	 6.7 	  & 	 7.88 	  & 	 8.1 	  & 	 8.5 	  & 	 -0.096 	  & 	 0.2 	  & 	 2 	  \\
 \object{2223$-$052} 	  & 	 3C 446 	  & 	 HPQ 	  & 	 22 	  & 	 1.404 	  & 	 G 	  & 	 14.31 	  & 	 16.0 	  & 	 16.470 	  & 	 16.5 	  & 	 3.6 	  & 	 -0.203 	  & 	 17.3 	  & 	 10 	  \\
 \object{2227$-$088} 	  & 	  	  & 	 HPQ 	  & 	 22 	  & 	 1.562 	  & 	 A 	  & 	 14.31 	  & 	 15.9 	  & 	 4.95 	  & 	 8.8 	  & 	 2.0 	  & 	 -0.137 	  & 	 6.9 	  & 	 12 	  \\
 \object{2230+114} 	  & 	 CTA 102 	  & 	 HPQ 	  & 	 37 	  & 	 1.037 	  & 	 E 	  & 	 14.28 	  & 	 15.6 	  & 	 15.51 	  & 	 15.5 	  & 	 3.7 	  & 	 -0.269 	  & 	 10.9 	  & 	 7 	  \\
 \object{2234+282} 	  & 	  	  & 	 HPQ 	  & 	 22 	  & 	 0.795 	  & 	 A 	  & 	 13.03 	  & 	 6.0 	  & 	 ... 	  & 	 ... 	  & 	 ... 	  & 	 -0.464 	  & 	 4.4 	  & 	 11 	  \\
 \object{2251+158} 	  & 	 3C 454.3 	  & 	 HPQ 	  & 	 37 	  & 	 0.859 	  & 	 E 	  & 	 15.26 	  & 	 33.2 	  & 	 14.86 	  & 	 19.9 	  & 	 1.3 	  & 	 -0.564 	  & 	 16.0 	  & 	 1 	  \\
 \object{2254+074} 	  & 	 PKS 2254+074 	  & 	 BLO 	  & 	 22 	  & 	 0.190 	  & 	 A 	  & 	 11.91 	  & 	 2.5 	  & 	 ... 	  & 	 ... 	  & 	 ... 	  & 	 ... 	  & 	 21.0 	  & 	 1 	  \\
\hline
\end{longtable}
\footnotesize{References: (1) \cite{angel80}; (2) \cite{impey90}; (3) \cite{impey91}; (4) \cite{impey00}; (5) \cite{jorstad07}; (6) \cite{lister00}; (7) \cite{moore84}; (8) \cite{punsly96}; (9) \cite{sluse05}; (10) \cite{smith85}; (11) \cite{stickel94}; (12) \cite{wills92}}.

}
}
\end{document}